\documentclass{JFM-FLM_Au}

\usepackage{amsmath}
\usepackage{amsfonts}
\usepackage{amssymb}

\usepackage{graphicx}
\usepackage{dcolumn}
\usepackage{bm}
\usepackage{bbm}
\usepackage{rotating}

\newcommand{\kn}{\mathrm{Kn}}

\usepackage{color}

\usepackage{enumerate,subfigure}

\lefttitle{T. Tomita, S. Taguchi and T. Tsuji}
\righttitle{Journal of Fluid Mechanics}

\title{Slow uniform flow of a rarefied gas \\past an infinitely thin circular disk}

\author{Takuma Tomita\aff{1}, Satoshi Taguchi\aff{1} \and Tetsuro Tsuji\aff{1}}

\affiliation{\aff{1}Graduate School of Informatics,
Kyoto University, Kyoto 606-8501, Japan
}

\corresau{Satoshi Taguchi, \email{taguchi.satoshi.5a@kyoto-u.ac.jp}}

\begin{document}
\maketitle

\begin{abstract}
The classical problem of steady rarefied gas flow past an infinitely thin circular disk is revisited, with particular emphasis on the gas behavior near the disk edge. The uniform flow is assumed to be perpendicular to the disk surface. An integral equation for the velocity distribution function, derived from the linearized Bhatnagar–Gross–Krook (BGK) model of the Boltzmann equation and subject to diffuse reflection boundary conditions, is solved numerically. The numerical method fully accounts for the discontinuity in the velocity distribution function that arises due to the presence of the edge. It is found that a kinetic boundary layer forms near the disk edge, extending over several mean free paths, and that its magnitude scales as $\mathrm{Kn}^{1/2}$ as the Knudsen number $\mathrm{Kn}$ (defined with respect to the disk radius) tends to zero. A thermal polarization effect, previously studied for spherical geometries, is also observed in the disk case, with a more pronounced manifestation near the edge that exhibits the same $\mathrm{Kn}^{1/2}$ scaling. The drag force acting on the disk is computed over a wide range of Knudsen numbers and shows good agreement with existing results for a hard-sphere gas and in the near-free-molecular regime.
\end{abstract}

\begin{keywords}
\end{keywords}


\section{\label{sec:introduction}Introduction}
Slow flows around small bodies provide fundamental insights into micro- and nanofluidics, optofluidics, and aerosol engineering. Classical Stokes-flow solutions remain valuable for deepening our understanding of active microswimmers and for developing new applications \citep{Piro_et_al_PhysRevRes_2024,Daddi_Moussa-Ider_PhysFluids_2020,Abdi_JMicrobioRobot_2021}. As the size of the suspended body decreases, non-equilibrium flow phenomena (or non-Navier-Stokes effects) become increasingly important. Typical examples in gaseous media include thermophoresis \citep{Takata-Sone_EJM1995,Takata_PHF09_thermophoresis,bosworth_ventura_ketsdever_gimelshein_2016,kalempa_sharipov_jfm_2020} and the motion of Janus particles \citep{Nosenko_etal_Janus_particles_in_plasma_PhysRevResearch_2020,Baier+Tiwari+Shrestha+Klar+Hardt_PhysRevFluids_2018}. 
In such cases, the mean free path of gas molecules becomes comparable to the characteristic size of the object, and the gas is regarded as being rarefied. While rarefied gas flows around spherical bodies have been thoroughly investigated across a broad range of Knudsen numbers \citep[see, e.g.,][]{Loyalka-PHF-1992,Takata-Sone-Aoki_PHF93,Taguchi+Suzuki_PhysRevFluids_2017,kalempa_sharipov_jfm_2020}, systematic studies for nonspherical geometries remain limited. Here, the Knudsen number is defined as the ratio of the molecular mean free path to a characteristic length scale, such as the sphere radius. The case of spheroidal particles has been studied in \citet{Aoki-Takata-Tomota_JFM-2014,Livi_et_al_PhysRevE_2022,Clercx_etal_EurJMechB_2024,Zhang_etal_PoF_2025}. In this study, we consider the rarefied gas flow past a circular disk of infinitesimal thickness. Despite its geometric simplicity, the edge configuration introduces a distinctive flow feature absent in smooth geometries, and its analysis is substantially more complex than that of a sphere, as described below.

Owing to its geometric simplicity, the rarefied gas flow around a circular disk is often regarded as a classical example of axisymmetric flow. Analytical expressions for the force acting on the disk have been derived in both the free-molecular flow limit \citep[][see also references therein]{Dahneke_1973,Ivanov-Yanshin_FD80} and in the near-free-molecular regime \citep{Sengers_etal_2014}. Numerical investigations have also been carried out in the high-speed flow regime \citep{Shakhov_1974,Chpoun_Elizarova_Graur_Lengrand_2005}. A simple plate geometry (with or without thickness) has also been used as a model to study the unsteady decay of a moving body in a free-molecular gas \citep{Aoki_Cavallaro_Marchioro_Pulvirenti_M2NA08,Aoki_Tsuji_Cavallaro_PhysRevE_2009}. However, most existing studies have focused on drag evaluation or high-speed flow behavior, and detailed investigations of the flow field structure in the low-speed regime, which is relevant to micro- and nanoscale applications, remain limited. In this context, the aim of the present study is twofold.

First, we aim to investigate the flow structure in the vicinity of the disk edge, where spatial variations in the flow variables occur on the scale of the molecular mean free path. Such edge effects have been studied primarily in the context of thermally induced flows \citep{Aoki-Sone-Masukawa95,Sone-Yoshimoto97,Taguchi-Aoki_JFM12,Ketsdever_Vacuum12} and their applications \citep[e.g.,][]{Taguchi-Aoki-PRE2015,Baier+Hardt+Shahabi+Roohi_PhysRevFluids_2017,Taguchi_tsuji_SciRep_2022,lotfian_roohi_JFM2019,Wang_etal_Microsystems_Nanoeng_2020}. However, much less is known about edge-induced effects in externally driven flows, despite their importance for understanding the behavior of particles in motion within a gas. In particular, we focus on the formation of a localized kinetic boundary layer, referred to as the edge layer, which is distinct from the conventional Knudsen layer \citep{Sone07} that forms over smooth surfaces.

The second objective of the present study is to demonstrate the feasibility of numerically analyzing a three-dimensional flow while accurately capturing discontinuities in the velocity distribution function (VDF). In the current problem, such discontinuities arise at the disk edge and propagate into the gas, and their precise representation is essential for correctly describing the flow field. 
One approach used in previous studies is a hybrid method that combines the finite-difference method with the method of characteristics. This method, originally introduced in \citet{Sugimoto-Sone92} and applied to two-dimensional thermal edge flows in \citet{Taguchi-Aoki_JFM12}, offers both accuracy and computational efficiency. However, its implementation becomes increasingly difficult as the number of independent variables increases, making it impractical for the present three-dimensional setting.
An alternative approach involves solving integral equations for macroscopic variables \citep[see Chap.~A.4.2 of][]{Sone07}. This method accurately handles VDF discontinuities and is memory-efficient. It has been successfully applied to rarefied flows in rectangular cavities \citep{Varoutis_Valougeorgis_Sharipov_2008,Hattori_2024}, where discontinuities arise at cavity corners. However, its implementation becomes considerably more involved in non-convex domains, since the integration domain, defined as the visible region from a given point, varies spatially and requires extensive problem-specific treatment. See, for example, its application to flows between non-coaxial cylinders \citep{Aoki-Sone-Yano89} or through cylinder arrays \citep{Taguchi-Charrier08}.

In this study, we adopt a more direct strategy by integrating the original equation along its characteristics. Since the VDF itself is treated as the unknown, this method requires more numerical integration. Nevertheless, the associated computational cost can be alleviated through the use of high-performance computing. The present method is motivated by \citet{taguchi_tsuji_kotera_jfm_2021} \citep[see also][]{Tsuji+Aoki_JCP2013}, where an unsteady rarefied gas flow involving a discontinuous velocity distribution function was analyzed by solving integral equations formulated for the VDF. 
We consider this approach a viable alternative to the finite-difference method for studying rarefied gas flows in complex geometries.

The remaining part of the paper is structured as follows. In Section \ref{sec:formulation}, we present the problem, providing its mathematical formulation based on the Bhatnagar--Gross--Krook (BGK) model \citep{Bhatnagar-Gross-Krook54,Welander54}. A detailed description of the discontinuity in the velocity distribution function is given in Section \ref{sec:discontinuity}. In Sec.~\ref{sec:numerical}, we derive an integral equation from the formulation in Sec.~\ref{sec:formulation} and outline the numerical procedure. Section~\ref{sec:freemolecular} discusses the case of a free molecular gas. Section~\ref{sec:numerical results} presents the numerical results, focusing on the behavior of the velocity distribution function and macroscopic quantities, followed by further discussions in Section \ref{sec:discussions}. Section \ref{sec:force} summarizes the result on the drag force acting on the disk. Finally, Section \ref{sec:conclusion} provides a brief summary of the findings.

\begin{figure}
    \centering
    \includegraphics{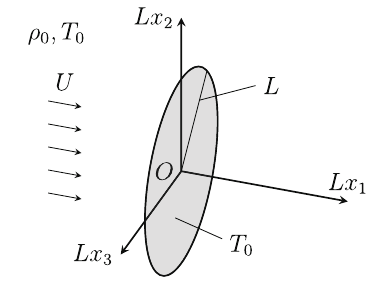}
    \caption{Problem: a flow past a circular disk.}
    \label{fig:problem}
\end{figure}

\section{\label{sec:formulation}Formulation}
\subsection{\label{subsec:problem}Problem}
Consider an infinitely thin circular disk (plate) with radius $L$ immersed in an ideal monatomic gas.
Let $Lx_i$ ($i = 1,2,3$) be a Cartesian coordinate system such that the origin $O$ is located at the center of the disk and the disk lies in the $x_2\,x_3$ plane with the $x_1$ axis perpendicular to the disk (see Fig.~\ref{fig:problem}). The corresponding position vector is denoted by $L\bm{x}$.
We assume that the disk is kept at a constant and uniform temperature $T_0$. Far from the disk, the gas is assumed to be in a uniform equilibrium state with velocity $(U,0,0)$, density $\rho_0$, temperature $T_0$, and pressure $p_0 = \rho_0 RT_0$, where $R$ denotes the gas constant per unit mass (i.e., the specific gas constant).
No external force is assumed to be present.
We investigate the steady behavior of the gas around the disk under the following assumptions.
\begin{enumerate}[(i)]
\item The behavior of the gas is described by the BGK model of the Boltzmann equation.
\item Gas molecules make diffuse reflections upon colliding with the surface of the disk.
\item The flow speed at infinity $U$ is much smaller than the thermal velocity, $(2RT_0)^{1/2}$. Consequently, the equations and boundary conditions can be linearized around the corresponding reference state at rest.
\end{enumerate}
Note that the diffuse reflection condition is adopted for numerical simplicity; more sophisticated boundary conditions can be used with additional computational effort.

\subsection{\label{subsec:formulation}Formulation}
We introduce our notation as follows: $(2RT_0)^{1/2} \zeta_i$ (or $(2RT_0)^{1/2} \bm{\zeta}$) is the molecular velocity, $\rho_0 (2RT_0)^{-3/2} (1 + \phi(\bm{x}, \bm{\zeta})) E$ is the velocity distribution function (VDF), $\rho_0 (1 + \omega(\bm{x}))$ is the density, $(2RT_0)^{1/2} u_i(\bm{x})$ is the flow velocity, $T_0 (1 + \tau(\bm{x}))$ is the temperature, $p_0 (1 + P(\bm{x}))$ is the pressure, and $p_0 (\delta_{ij} + P_{ij}(\bm{x}))$ is the stress tensor.
Here, $E = \pi^{-3/2} \exp (-\bm{\zeta}^2)$ and $\delta_{ij}$ is the Kronecker delta.

The linearized BGK equation, the diffuse reflection condition on the disk, and the condition at infinity for the present steady problem read
\begin{subequations}
    \label{eq:basic_eq_Cartesian}
\begin{align}
    \label{eq:LBGK_eq}
    &\zeta_i \frac{\partial \phi}{\partial x_i} = \frac{1}{\kappa} \mathcal{L}(\phi), \\
    \label{eq:def_L}
    &\mathcal{L}(\phi) = -\phi + \omega + 2\zeta_i u_i + \left(\zeta_j^2 - \frac{3}{2} \right) \tau, \\
    \label{eq:out}
    &\omega = \int \phi E \mathrm{d}\bm{\zeta}, \quad
     u_i = \int \zeta_i \phi E \mathrm{d}\bm{\zeta}, \quad
     \tau = \frac{2}{3} \int (\zeta_j^2 - \frac{3}{2}) \phi E \mathrm{d}\bm{\zeta}, \\
    \label{eq:b.c._disk}
    & \text{b.c. } 
    \phi = 2\sqrt{\pi} \int_{-\infty}^{\infty}
     \int_{-\infty}^{\infty} \int_{0}^{\mp \infty} \zeta_1 \phi E  \mathrm{d}\zeta_1 \mathrm{d}\zeta_2 \mathrm{d}\zeta_3
     \quad
     \text{for} \quad \zeta_1 \gtrless 0 \quad
     (x_1 = \pm 0, \,\, x_2^2 + x_3^2 < 1), \\
    \label{eq:b.c._faraway}
    & \text{b.c. } 
    \phi \to 2\zeta_1 u_{\infty} \quad (|\bm{x}| \to \infty).
\end{align}
\end{subequations}
Here, $\kappa$ in Eq.~\eqref{eq:LBGK_eq} is the parameter defined by
\begin{align}
\label{eq:def_kappa}
    \kappa = \frac{\sqrt{\pi}}{2} \frac{\ell_0}{L} = \frac{\sqrt{\pi}}{2} \mathrm{Kn},
\end{align}
where $\ell_0$ denotes the molecular mean free path at the reference equilibrium state and $\mathrm{Kn} (= \ell_0/L$) the Knudsen number.
Note that the BGK model has $\ell_0$ which is calculated as $\ell_0 = (2/\sqrt{\pi})(2RT_0)^{1/2}/A_c \rho_0$, where $A_c$ is a constant.
In Eq.~\eqref{eq:b.c._faraway}, $u_{\infty}$ denotes the dimensionless flow velocity at infinity, given by $u_{\infty} = (2RT_0)^{-1/2} U$.
The pressure and stress tensor are expressed in terms of $\phi$ through the following integrals:
\begin{align}
\label{eq:ppq}
    P = \frac{2}{3} \int \zeta_j^2 \phi E \mathrm{d}\bm{\zeta} = \omega + \tau, \quad
    P_{ij} = 2 \int \zeta_i \zeta_j \phi E \mathrm{d}\bm{\zeta}.
\end{align}

\subsection{\label{subsec:transformation}Coordinate transformation}

Let $(Lx, Lr, \theta)$ be the cylindrical coordinates; $x_1 = x$, $x_2 = r \cos \theta$, and $x_3 = r \sin \theta$. The components of molecular velocity in these coordinates are denoted as $(\zeta_x, \zeta_r, \zeta_{\theta})$, where $\zeta_1 = \zeta_x$, $\zeta_2 = \zeta_r \cos \theta - \zeta_{\theta} \sin \theta$, and $\zeta_3 = \zeta_r \sin \theta + \zeta_{\theta} \cos \theta$. The same convention applies to other vectors or tensors, such as $u_x$, $P_{xr}$, etc. Furthermore, the local polar coordinates $(\zeta, \theta_{\zeta}, \varphi_{\zeta})$ are introduced to express the molecular velocity components as follows: $\zeta_x = \zeta \cos \theta_{\zeta}$, $\zeta_r = \zeta \sin \theta_{\zeta} \cos \varphi_{\zeta}$, and $\zeta_{\theta} = \zeta \sin \theta_{\zeta} \sin \varphi_{\zeta}$.
The velocity distribution function in the new coordinate system is expressed as $\phi_C=\phi_C(x,r,\theta,\zeta,\theta_{\zeta},\varphi_{\zeta})$. If the flow is assumed to be axisymmetric, $\phi_C$ is independent of $\theta$.
It is straightforward to verify that $\phi_C$ satisfies the following symmetry properties (cf. Eq.~\eqref{eq:basic_eq_cylind}  below):
\begin{subequations}
\label{eq:phiC_symm_propeties}
\begin{align}
    \label{eq:phiC_even}
    &\phi_C(x,r,\zeta,\theta_{\zeta},-\varphi_{\zeta}) = \phi_C(x,r,\zeta,\theta_{\zeta},\varphi_{\zeta}), \\
    \label{eq:phiC_skew}
    &\phi_C(-x,r,\zeta,\pi-\theta_{\zeta},\varphi_{\zeta}) = -\phi_C(x,r,\zeta,\theta_{\zeta},\varphi_{\zeta}).
\end{align}
\end{subequations}
Equation~\eqref{eq:phiC_even} indicates that $\phi_C$ is even with respect to $\varphi_{\zeta}$, allowing the range of $\varphi_{\zeta}$ to be restricted to $0 \le \varphi_{\zeta} \le \pi$.
The range $- \infty < x < \infty$ is restricted to $x > 0$ by the condition
\begin{align}
    \label{eq:b.c._skew}    &\phi_C(x,r,\zeta,\theta_{\zeta},\varphi_{\zeta}) = -\phi_C(x,r,\zeta,\pi-\theta_{\zeta},\varphi_{\zeta})
    \quad
    (x=0_+, \,\, r>1, \,\, 0 \leq \theta_{\zeta} \leq \pi/2),
\end{align}
derived from \eqref{eq:phiC_skew}.
Here and hereafter, $0_+$ (and $0_-$) means $\lim_{\epsilon \downarrow 0}0+\epsilon$ (and $\lim_{\epsilon \downarrow 0}0-\epsilon$).
The value of $\phi_C$ for $x<0$ can be determined from its value for $x>0$ using \eqref{eq:phiC_skew}.

Now, we present the equations governing $\phi_C$.
Let $D$ denote the domain defined as
\begin{align}
    \label{eq:def_D}
    &D = \left\{ (x,r,\zeta,\theta_{\zeta},\varphi_{\zeta}) \in \mathbb{R}^5 \mid 
                x > 0, \, r \geq 0,  \, \zeta \geq 0, \, 
                0 \leq \theta_{\zeta} \leq \pi, \, 0 \leq \varphi_{\zeta} \leq \pi  \right\}.
\end{align}
The equation and boundary conditions that $\phi_C$ satisfies are given as follows:
\begin{subequations}
\label{eq:basic_eq_cylind}
\begin{align}
    \label{eq:LBGK_cylind}
    &\zeta \cos \theta_{\zeta} \frac{\partial \phi_C}{\partial x}
    + \zeta \sin \theta_{\zeta} \cos \varphi_{\zeta} \frac{\partial \phi_C}{\partial r}
    - \frac{\zeta \sin \theta_{\zeta} \sin \varphi_{\zeta}}{r} \frac{\partial \phi_C}{\partial \varphi_{\zeta}}
    = \frac{1}{\kappa} \mathcal{L}_1(\phi_C), \quad \text{in $D$},\\
    \label{eq:b.c._disk_cylind}
    &\text{b.c. } \phi_C = \sigma_{\mathrm{w}} \quad
    (x = 0_+, \,\, 0 \leq r < 1, \,\, 0 \leq \theta_{\zeta} \leq \pi/2), \\
    \label{eq:b.c._outofdisk_cylind}
    &\text{b.c. } \phi_C(x, r, \zeta, \theta_{\zeta}, \varphi_{\zeta})
    = -\phi_C(x, r, \zeta, \pi-\theta_{\zeta}, \varphi_{\zeta})
    \quad (x=0_+, \,\, r>1, \,\, 0 \leq \theta_{\zeta} \leq \pi/2), \\
    \label{eq:b.c._faraway_cylind}
    &\text{b.c. } \phi_C \to 2 \zeta u_\infty \cos \theta_{\zeta} \quad (x^2+r^2 \to \infty),
\end{align}
\end{subequations}
where the operator $\mathcal{L}_1$ is defined as
\begin{subequations}
\begin{align}
    \label{eq:def_L1_cylind}
    \mathcal{L}_1(\phi_C) & = -\phi_C + \omega + 2\zeta u_x \cos \theta_{\zeta}
     + 2 \zeta u_r \sin \theta_{\zeta} \cos \varphi_{\zeta} + (\zeta^2 - \frac{3}{2}) \tau, \\
    \label{eq:omega_cylind}
    \omega & = 2 \int_{0}^{\pi} \int_{0}^{\pi} \int_{0}^{\infty} 
     \zeta^2 \sin \theta_{\zeta} \phi_C E
     \mathrm{d}\zeta \mathrm{d}\theta_{\zeta} \mathrm{d}\varphi_{\zeta}, \\
    \label{eq:ux_cylind}
    u_x & = 2 \int_{0}^{\pi} \int_{0}^{\pi} \int_{0}^{\infty}
     \zeta^3 \sin \theta_{\zeta} \cos \theta_{\zeta} \phi_C E
     \mathrm{d}\zeta \mathrm{d}\theta_{\zeta} \mathrm{d}\varphi_{\zeta}, \\
    \label{eq:ur_cylind}
    u_r & = 2 \int_{0}^{\pi} \int_{0}^{\pi} \int_{0}^{\infty}
     \zeta^3 \sin^2 \theta_{\zeta} \cos \varphi_{\zeta} \phi_C E
     \mathrm{d}\zeta \mathrm{d}\theta_{\zeta} \mathrm{d}\varphi_{\zeta}, \\
    \label{eq:tau_cylind}
    \tau & = \frac{4}{3} \int_{0}^{\pi} \int_{0}^{\pi} \int_{0}^{\infty}
     \zeta^2 (\zeta^2 - \frac{3}{2}) \sin \theta_{\zeta} \phi_C E
     \mathrm{d}\zeta \mathrm{d}\theta_{\zeta} \mathrm{d}\varphi_{\zeta},
\end{align}
\end{subequations}
and $\sigma_{\mathrm{w}} = \sigma_{\mathrm{w}}(r)$ is given by
\begin{align}
    \label{eq:def_sigmaw}
    &\sigma_{\mathrm{w}}(r)
    = -4\sqrt{\pi} \int_{0}^{\pi} \int_{\pi/2}^{\pi} \int_{0}^{\infty} \zeta^3 
       \sin \theta_{\zeta} \cos \theta_{\zeta}
       \phi_C(0_{+}, r, \zeta, \theta_{\zeta}, \varphi_{\zeta}) 
       E \mathrm{d}\zeta \mathrm{d}\theta_{\zeta} \mathrm{d}\varphi_{\zeta}, \notag \\ &\hspace{10cm} \quad 0 \le r < 1.
\end{align}
The macroscopic quantities are expressed in terms of $\phi_C$ as follows:
\begin{subequations}
\label{eq:ppq_cylind}
\begin{align}
\label{eq:pxx_cylind}
    &P_{xx} = 4 \int_{0}^{\pi} \int_{0}^{\pi} \int_{0}^{\infty}
             \zeta^4 \sin \theta_{\zeta} \cos^2 \theta_{\zeta} \phi_C E
             \mathrm{d}\zeta \mathrm{d}\theta_{\zeta} \mathrm{d}\varphi_{\zeta}, \\
\label{eq:prr_cylind}
    &P_{rr} = 4 \int_{0}^{\pi} \int_{0}^{\pi} \int_{0}^{\infty}
             \zeta^4 \sin^3 \theta_{\zeta} \cos^2 \varphi_{\zeta} \phi_C E
             \mathrm{d}\zeta \mathrm{d}\theta_{\zeta} \mathrm{d}\varphi_{\zeta}, \\
\label{eq:ptt_cylind}
    &P_{\theta\theta} = 4 \int_{0}^{\pi} \int_{0}^{\pi} \int_{0}^{\infty}
             \zeta^4 \sin^3 \theta_{\zeta} \sin^2 \varphi_{\zeta} \phi_C E
             \mathrm{d}\zeta \mathrm{d}\theta_{\zeta} \mathrm{d}\varphi_{\zeta}, \\
\label{eq:pxr_cylind}
    &P_{xr} = 4 \int_{0}^{\pi} \int_{0}^{\pi} \int_{0}^{\infty}
             \zeta^4 \sin^2 \theta_{\zeta} \cos \theta_{\zeta} \cos \varphi_{\zeta} \phi_C E
             \mathrm{d}\zeta \mathrm{d}\theta_{\zeta} \mathrm{d}\varphi_{\zeta}, \\
\label{eq:macro_cylind}
    &u_{\theta} = P_{x \theta} = P_{r \theta} = 0.
\end{align}
\end{subequations}

The force acting on the disk is directed along the $x_1$ (or $x$) axis due to symmetry. Denoting the $x_1$ component of the force by $F$, it is expressed as
\begin{subequations}
\label{eq:drag}
\begin{align}
    \label{eq:drag_force}
    &F = p_0 L^2 (2RT_0)^{-1/2} U h_D, \\
    \label{eq:def_h_D}
    &h_D = -4\pi \int_{0}^{1} \frac{P_{xx}(x=0_+,r)}{u_{\infty}} r \mathrm{d}r.
\end{align}
\end{subequations}
Here, $h_D$ is a function of $\kappa$ (or the Knudsen number), i.e., $h_D=h_D(\kappa)$, and it characterizes the effect of $\kappa$ on the drag force.
One of the key objectives of this study is to understand how the Knudsen number $\kappa$ influences $h_D$.

\section{\label{sec:discontinuity}Discontinuity of the velocity distribution function}


The left-hand side of the BGK equation \eqref{eq:LBGK_eq}
represents the rate of change of $\phi$ along the characteristics, which correspond to straight lines in the $\bm{x}$ space.
This implies that the value of $\phi$ at $\bm{x}$ for a given $\bm{\zeta}$ is determined by integrating the right-hand side along the half-line $\tilde{\bm{x}}(s) = \bm{x} -(\bm{\zeta}/\zeta) s$, where $s (\ge 0)$ represents the distance from $\bm{x}$. 
Thus, depending on whether the half line (backward characteristics) intersects the disk or not, the value of $\phi(\bm{x},\bm{\zeta})$ is influenced by the diffuse reflection condition on the disk or by the integration over contributions from infinity. Consequently, $\phi(\bm{x},\bm{\zeta})$ undergoes an abrupt change at $\bm{\zeta}$ where the half-line transitions from intersecting to not intersecting the disk. The discontinuity thus created on the edge propagates through the gas along the characteristics. 
In summary, the discontinuity jump occurs for those $\bm{\zeta}$ such that $-\bm{\zeta}$ lies on the conical surface with its apex at $\bm{x}$ and its base coinciding with the disk.


\begin{figure*}
\begin{center}
\subfigure[]{\includegraphics[scale=1.1]{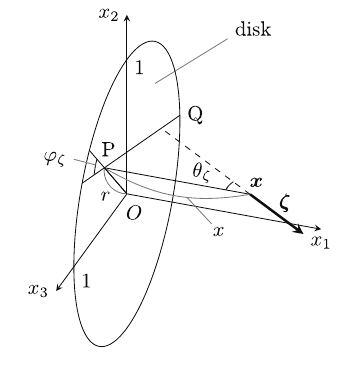}}
\qquad
\subfigure[]{\includegraphics[scale=1.1]{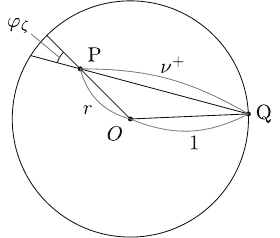}}
\caption{(a) Backward characteristics (dashed line) from the point $\bm{x}$ in the direction of $-\bm{\zeta}$ in the case of $0 \leq r < 1$. The thick solid arrow indicates the molecular velocity $\bm{\zeta}$.
(b) shows a projected view of from the positive side of the $x_1$ axis.}
\label{fig:trajectory_0r1}
\end{center}
\end{figure*}

To analyze the precise location of the discontinuity in the domain $D$, we first consider the case $0 \le r < 1$ (see Fig.~\ref{fig:trajectory_0r1}). The point $\bm{x}$ is projected onto the plane $x_1=0$, referred to as P.
Next, we draw a line from P with an angle $\varphi_{\zeta} \ge 0$, which intersects the perimeter of the disk at a point Q. Tracing the characteristics from $\bm{x}$ in the $-\bm{\zeta}$ direction, the projection onto the plane $x_1 = 0$ moves along the line PQ from P towards Q.
Therefore, the condition for the backward characteristic line to hit the disk can be expressed as
\begin{align}
    \label{eq:enc_cond_0r1}
    0 \leq \theta_{\zeta} \leq \arctan \left(\frac{\nu^{+}}{x}\right).
\end{align}
Here, $\nu^{+}$ denotes the length of PQ, defined as
\begin{align}
\label{eq:def_nup}
    \nu^{+} = \nu^{+}(r,\varphi_{\zeta}) := r \cos \varphi_{\zeta} + \sqrt{1 - r^2 \sin^2 \varphi_{\zeta}}.
\end{align}

\begin{figure*}
\begin{center}
\subfigure[]{\includegraphics[scale=1.1]{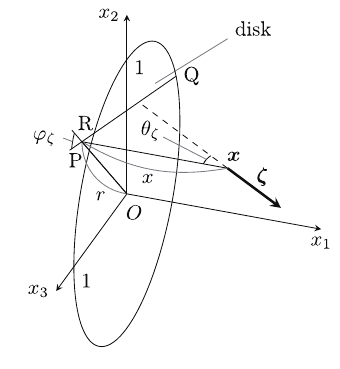}}
\qquad
\subfigure[]{\includegraphics[scale=1.1]{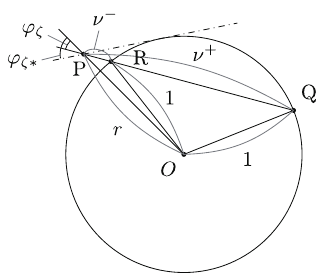}}
\caption{(a) Backward characteristics (dashed line) from the point $\bm{x}$ in the direction of $-\bm{\zeta}$ in the case of $r > 1$. See the caption of Fig.~\ref{fig:trajectory_0r1}.}
\label{fig:trajectory_r1}
\end{center}
\end{figure*}
Next, we consider the case $r > 1$ (see Fig.~\ref{fig:trajectory_r1}).
If we draw a line from P that intersects the disk perimeter at two points, Q and R (where R is closer to P than Q), the angle $\varphi_{\zeta}$ between the line PQ and PO must be in the range
\begin{align}
    \label{eq:enc_cond_r1_phi}
    \varphi_{\zeta} < \varphi_{\zeta *},
\end{align}
where
\begin{align}
    \label{eq:def_pzast}
    \varphi_{\zeta *} = \varphi_{\zeta *}(r) = \arcsin(1/r), \quad 0 \le  \varphi_{\zeta *} \le \pi.
\end{align}
If this is the case, the backward characteristic, when projected onto the plane $x_1 = 0$, can be traced along the line segment PQ from P towards Q.
Thus, the condition for the backward characteristic line to hit the disk can be expressed as
\begin{align}
    \label{eq:enc_cond_1r}
    \arctan \left(\frac{\nu^{-}}{x}\right) \leq \theta_{\zeta} \leq \arctan \left(\frac{\nu^{+}}{x} \right).
\end{align}
Here, $\nu^+$ represents the length of the segment PQ, as defined in \eqref{eq:def_nup}, and $\nu^-$ denotes the length of the segment PR, given by
\begin{align}
    \label{eq:def_num}
    \nu^{-} = \nu^{-}(r,\varphi_{\zeta}) := r \cos \varphi_{\zeta} - \sqrt{1 - r^2 \sin^2 \varphi_{\zeta}}.
\end{align}

Based on the above discussions, the condition for the backward characteristics to intersect the disk is summarized as follows. Let $\widetilde{D}$ be the domain defined by $(x,r,\theta_{\zeta},\varphi_{\zeta}) \in \mathbb{R}^+ \times \mathbb{R}^+ \times [0,\pi] \times [0,\pi]$.
Then, the condition is expressed as $(x,r,\theta_{\zeta},\varphi_{\zeta}) \in \Omega \subset \widetilde{D}$,
where
\begin{subequations}
\begin{align}
    \label{eq:def_Omega}
    & \Omega = \Omega_1 \cup \Omega_2, \\
    & \Omega_1 = \left\{ (x,r,\theta_{\zeta},\varphi_{\zeta}) \in \widetilde{D} \mid
                       0 \leq r < 1, \,\, 
                       0 \leq \theta_{\zeta} \le \theta_{\zeta*}^{+}(x,r,\varphi_{\zeta}) \right\},
                       \\
    & \Omega_2 = \left\{ (x,r,\theta_{\zeta},\varphi_{\zeta}) \in \widetilde{D} \mid 
                       r \geq 1, \,\, \theta_{\zeta*}^{-}(x,r,\varphi_{\zeta}) \le \theta_{\zeta} \le \theta_{\zeta*}^{+}(x,r,\varphi_{\zeta}), \,\,
                       0 \leq \varphi_\zeta < \varphi_{\zeta*} \right\}.
                       \label{eq:def_Omega_2}
\end{align}
\end{subequations}
Here, the following notation has been introduced:
\begin{align}
    \label{eq:def_tzastpm}
    \theta_{\zeta*}^{\pm} = \theta_{\zeta*}^{\pm}(x,r,\varphi_{\zeta})
                         := \arctan \left( \frac{\nu^{\pm}(r,\varphi_{\zeta})}{x} \right).
\end{align}
The limiting case $r \to 1$ has been included in the definition of $\Omega_2$.
The position of the discontinuity in $\widetilde{D}$ is determined by the boundary $\partial \Omega$ of $\Omega$, which forms
a surface in the four-dimensional space $(x,r,\theta_{\zeta},\varphi_{\zeta})$.
\begin{figure*}
\begin{center}
%
\subfigure[]{\includegraphics[scale=0.98]{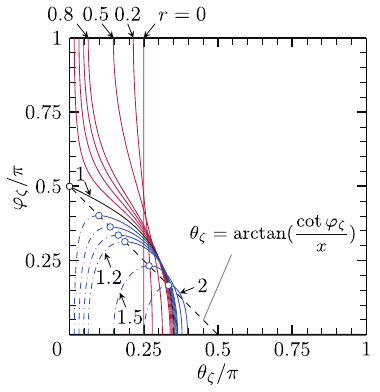}}
\qquad
\subfigure[]{\includegraphics[scale=0.98]{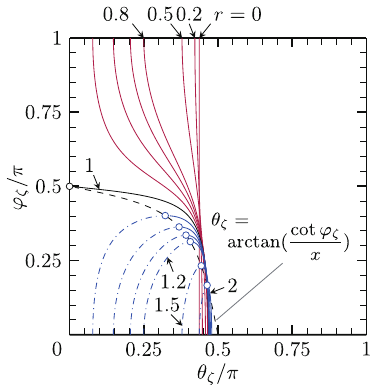}}
\caption{Cross sections of the boundary $\partial \Omega$ in the $\theta_\zeta\,\varphi_\zeta$ plane, where the VDF is discontinuous, for various values of $r$ in the cases of (a) $x=1$ and (b) $x=0.2$.
For a given $x$, the solid red curves represent $\theta_\zeta = \theta_{\zeta *}^+(x,r,\varphi_\zeta)$ as a function of $\varphi_\zeta$ ($r<1$); the solid (dash-dotted) blue curves represent $\theta_\zeta = \theta_{\zeta *}^+(x,r,\varphi_\zeta)$ ($\theta_\zeta = \theta_{\zeta *}^-(x,r,\varphi_\zeta)$) as a function of $\varphi_\zeta$ ($r>1$); the solid black curves represent $\theta_\zeta = \theta_{\zeta *}^+(x,r,\varphi_\zeta)$ as a function of $\varphi_\zeta$ ($r=1$).
The values of $r \in [0.8, 1.2]$ not shown in the panels are $r=0.8+0.05 m$ ($m=0,1,\ldots,8$).
When $r > 1$, the curve $\theta_{\zeta} = \theta_{\zeta *}^{+}$ (solid blue curves) and $\theta_{\zeta} = \theta_{\zeta *}^{-}$ (dash-dotted blue curves) are joined at $\varphi_\zeta = \varphi_{\zeta *}$ indicated by open circles.
The black dashed line indicates 
$\theta_\zeta = \text{arctan}(\cot \varphi_\zeta/x)$,
which gives the trajectory of $\varphi_\zeta = \varphi_{\zeta *}(r)$ for $r \ge 1$.}
\label{fig:discontinuity}
\end{center}
\end{figure*}

In Fig.~\ref{fig:discontinuity}, we show typical cross sections of $\partial \Omega$ in the $\theta_{\zeta}\,\varphi_{\zeta}$ plane for various values of $r$ in the cases of $x = 1$ [(a)] and $x=0.2$ [(b)]. 
In the panels, the solid red curves represent $\theta_\zeta = \theta_{\zeta *}^+(x,r,\varphi_\zeta)$ as a function of $\varphi_\zeta$ for given $x$ and $r<1$; the solid (dash-dotted) blue curves represent $\theta_\zeta = \theta_{\zeta *}^+(x,r,\varphi_\zeta)$ ($\theta_\zeta = \theta_{\zeta *}^-(x,r,\varphi_\zeta)$) 
for $r>1$; the solid black curves represent $\theta_\zeta = \theta_{\zeta *}^+(x,r,\varphi_\zeta)$ 
for $r=1$. In all cases, $\theta_\zeta$ is used as the abscissa.
The velocity distribution function $\phi_C$ exhibits a jump discontinuity along these curves.
For $r < 1$ (red curves), the function $\theta_{\zeta *}^+(x,r,\varphi_\zeta)$ decreases monotonically for $\varphi_\zeta \in [0,\pi]$.
For $r>1$ (blue curves), the function $\theta_{\zeta *}^+(x,r,\varphi_\zeta)$ ($\theta_{\zeta *}^-(x,r,\varphi_\zeta)$) decreases (increases) monotonically for $\varphi_\zeta \in [0,\varphi_{\zeta *}(r)]$. These two curves meet at the point $\varphi_\zeta = \varphi_{\zeta *}(r)$, marked by open circles in the panels. The change in behavior between $r < 1$ and $r > 1$ is clearly related to the number of intersections of the backward characteristic with the disk projected on the plane $x_1=0$, which varies depending on whether the point P lies inside or outside the disk perimeter, as discussed earlier (see Figs.~\ref{fig:trajectory_0r1} and \ref{fig:trajectory_r1}). Additionally, the trajectory of $\varphi_{\zeta *}(r)$ ($r \ge 1$) is given by $\theta_\zeta = \text{arctan}(\cot \varphi_\zeta/x)$ and is represented by the dashed line in the figure.

\section{\label{sec:numerical}Numerical analysis}

As discussed in the preceding section, one of the critical aspects of the present problem is the tip-induced discontinuity propagating from the edge. The precise location of this discontinuity in the four-dimensional space $(x,r,\theta_{\zeta},\varphi_{\zeta})$ is determined by complex equations. Handling such discontinuities using a finite-difference method is incredibly challenging. Instead, our approach relies on an integral equation formulation, as proposed in \citet{taguchi_tsuji_kotera_jfm_2021,Tsuji+Aoki_JCP2013}. 

\begin{figure*}
\begin{center}
\subfigure[]{\includegraphics[scale=1.2]{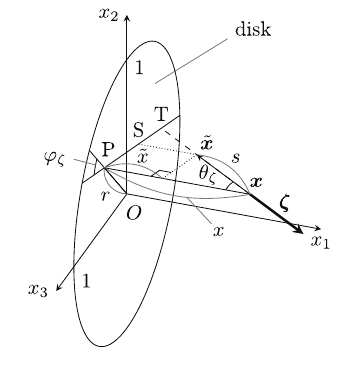}}
\qquad
\subfigure[]{\includegraphics[scale=1.2]{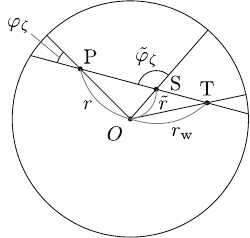}}
\caption{The geometrical interpretations of $\tilde{x}$, $\tilde{r}$, $\tilde{\varphi}_{\zeta}$, and $r_{\mathrm{w}}$ [(a)] and a view from the positive side of the $x_1$ axis [(b)]. Suppose that we move along the characteristics from $\bm{x}$ to $\tilde{\bm{x}} = \bm{x} - \bm{\ell} s$ for a given $\bm{\ell}=\bm{\zeta}/\zeta$.
Then, the cylindrical coordinates $(x,r)$ of $\bm{x}$ change to $(\tilde{x},\tilde{r})$ at $\tilde{\bm{x}}$. Furthermore, at $\tilde{\bm{x}}$, the azimuth angle $\varphi_{\zeta}$ of $\bm{\zeta}$ changes to $\tilde{\varphi}_{\zeta}$.
If we project the trajectory onto the plane $x_1=0$ and call the resulting segment PS,
the length of the segment OS gives $\tilde{r}$, and the angle between the two lines SP and OS gives $\tilde{\varphi}_{\zeta}$.
In the case where $(x,r,\theta_{\zeta},\varphi_{\zeta}) \in \Omega$, if the intersection of the characteristic with the disk is denoted by T, the length of the segment OT gives $r_{\mathrm{w}}$.}
\label{fig:traceback}
\end{center}
\end{figure*}

\subsection{\label{subsec:integralequation}Integral equations}
We begin with the case $(x,r,\theta_{\zeta},\varphi_{\zeta}) \in \Omega$; the case $(x,r,\theta_{\zeta},\varphi_{\zeta}) \notin \Omega$ will be discussed later.
Given $\bm{x}$ and $\bm{\zeta}$, the points in the backward characteristics can be expressed by $\tilde{\bm{x}}(s) = \bm{x} - \bm{\ell} s$, where $\bm{\ell} = \bm{\zeta}/\zeta$ and $s$ is the distance from $\bm{x}$. Note that the distance from $\bm{x}$ to the disk along the characteristics is given by \begin{align} \label{eq:def_sw} s_{\mathrm{w}} = s_{\mathrm{w}}(x,\theta_{\zeta}) := \frac{x}{\cos \theta_{\zeta}}, \end{align} and $s$ is treated as a parameter in the range $[0,s_\mathrm{w})$.
Integrating the BGK equation \eqref{eq:LBGK_eq}, multiplied by $(1/\kappa\zeta) \exp(-s/\kappa\zeta)$, over the range from $s=0$ to $s=s_{\mathrm{w}}$, we obtain an integral equation for $\phi_C$:\begin{align}
    \label{eq:inteq_enc}
    \phi_C(x,r,\zeta,\theta_{\zeta},\varphi_{\zeta})
    = \exp( -\frac{s_{\mathrm{w}}}{\kappa\zeta} ) \sigma_{\mathrm{w}}(r_{\mathrm{w}})
      + \frac{1}{\kappa\zeta} \int_{0}^{s_{\mathrm{w}}} \exp (-\frac{s}{\kappa\zeta})
        G(\tilde{x}(s), \tilde{r}(s), \zeta, \theta_{\zeta}, \tilde{\varphi}_{\zeta}(s)) \mathrm{d}s,
\end{align}
where
\begin{subequations}
\label{eq:inteq_remark}
\begin{align}
    \label{eq:def_rast}
    &r_{\mathrm{w}} = r_{\mathrm{w}}(x,r,\zeta,\theta_{\zeta},\varphi_{\zeta})
     := \sqrt{r^2 + x^2 \tan^2 \theta_{\zeta} - 2rx \tan \theta_{\zeta} \cos \varphi_{\zeta}}, \\
    \label{eq:def_G}
    &G(\tilde{x},\tilde{r},\zeta, \theta_{\zeta},\tilde{\varphi}_{\zeta})
     = \omega(\tilde{x},\tilde{r}) + 2\zeta u_x(\tilde{x},\tilde{r})\cos \theta_{\zeta}
       + 2 \zeta u_r(\tilde{x},\tilde{r}) \sin \theta_{\zeta} \cos \tilde{\varphi}_{\zeta}
       + (\zeta^2 - \frac{3}{2}) \tau(\tilde{x},\tilde{r}), \\
    \label{eq:def_tilx}
    &\tilde{x}(s) = \tilde{x}(s;x,\theta_{\zeta}) := x - s\cos \theta_{\zeta}, \\
    \label{eq:def_tilr}
    &\tilde{r}(s) = \tilde{r}(s;r,\theta_{\zeta},\varphi_{\zeta})
     := \sqrt{r^2 + s^2 \sin^2 \theta_{\zeta} - 2rs \sin \theta_{\zeta} \cos \varphi_{\zeta}}, \\
    \label{eq:def_tilphi}
    &\tilde{\varphi}_{\zeta}(s) = \tilde{\varphi}_{\zeta}(s;r,\theta_{\zeta},\varphi_{\zeta})
     := \begin{cases} \displaystyle \arcsin \left( \frac{r\sin \varphi_{\zeta}}{\tilde{r}(s)} \right), 
                      &\displaystyle \text{if } s \leq \frac{r\cos\varphi_{\zeta}}{\sin\theta_{\zeta}}, \\
                      \displaystyle \pi - \arcsin \left( \frac{r\sin \varphi_{\zeta}}{\tilde{r}(s)} \right),
                      &\displaystyle \text{if } s > \frac{r\cos\varphi_{\zeta}}{\sin\theta_{\zeta}}. \end{cases}
\end{align}
\end{subequations}
The geometrical meanings of $\tilde{x}$, $\tilde{r}$, $\tilde{\varphi}_{\zeta}$, and $r_{\mathrm{w}}$ are illustrated in Fig.~\ref{fig:traceback}.

Next, we consider the case $(x,r,\theta_{\zeta},\varphi_{\zeta}) \notin \Omega$, where the backward characteristics extend to infinity without intersecting the disk.
A similar analysis to that in the previous case leads to the following equation:
\begin{align}
    \label{eq:inteq_notenc}
    \phi_C(x,r,\zeta,\theta_{\zeta},\varphi_{\zeta})
    = \frac{1}{\kappa\zeta} \int_{0}^{\infty} \exp (-\frac{s}{\kappa\zeta}) 
      G(\tilde{x}(s), \tilde{r}(s), \zeta, \theta_{\zeta}, \tilde{\varphi}_{\zeta}(s)) \mathrm{d}s,
\end{align}
where $G$, $\tilde{x}(s)$, $\tilde{r}(s)$, and $\tilde{\varphi}_{\zeta}(s)$ are as defined in \eqref{eq:def_G}--\eqref{eq:def_tilphi}.
Since the backward characteristics extend to infinity, no boundary term is present in \eqref{eq:inteq_notenc}. The boundary condition \eqref{eq:b.c._faraway_cylind} (or \eqref{eq:b.c._faraway}) is implicitly incorporated into the right-hand side of the equation through the macroscopic quantities included in $G$.

To summarize, the integral equations derived for the two cases $(x,r,\theta_{\zeta},\varphi_{\zeta}) \in \Omega$ and $(x,r,\theta_{\zeta},\varphi_{\zeta}) \notin \Omega$ can be unified into a single equation:
\begin{align}
    \label{eq:inteq_unfied}
    \phi_C(x,r,\zeta,\theta_{\zeta},\varphi_{\zeta})
    & = \exp( -\frac{s_{\mathrm{w}}}{\kappa\zeta} ) \sigma_{\mathrm{w}}(r_{\mathrm{w}}) \mathbbm{1}_\Omega
    \nonumber \\
    &  + \frac{1}{\kappa\zeta} \int_{0}^{s_*} \exp (-\frac{s}{\kappa\zeta})
        G(\tilde{x}(s), \tilde{r}(s), \zeta, \theta_{\zeta}, \tilde{\varphi}_{\zeta}(s)) \mathrm{d}s,
\end{align}
where 
$\mathbbm{1}_\Omega$ is the characteristic function of $\Omega$ and
\begin{align} \label{eq:def_sast}
    s_* = s_*(x,r,\theta_\zeta,\varphi_\zeta) = \begin{cases}
        s_{\mathrm{w}}(x,\theta_\zeta), \quad (x,r,\theta_\zeta,\varphi_\zeta) \in \Omega, \\
        \infty, \quad (x,r,\theta_\zeta,\varphi_\zeta) \notin \Omega.
    \end{cases}
\end{align}
The discontinuity of $\phi_C$ is reflected in this expression through the dependence on $\mathbbm{1}_\Omega$ and $s_*$.

\subsection{\label{subsec:outline}Outline of the numerical computations}

For numerical computation, we introduce the following parametric expressions (oblate spheroid coordinates) for $x$ and $r$:
\begin{align} \label{eq:oblate}
 x = \sinh \xi \cos \eta, \quad r = \cosh \xi \sin \eta, \quad 0 \le \xi < \infty, \quad 0 \le \eta \le \pi/2.
\end{align}
We then restrict the range of $\xi$ and that of $\zeta$ to $0 \le \xi \le \xi_{\mathrm{max}}$ and $0 \le \zeta \le \zeta_{\mathrm{max}}$, respectively, where $\xi_{\mathrm{max}}$ and $\zeta_{\mathrm{max}}$ are sufficiently large values chosen to approximate the infinite domain.
Note that \eqref{eq:oblate} defines a meridian plane in the oblate spheroid coordinate system, with the $x$ axis being the axis of rotation.
To discretize the domain, we introduce lattice points for $(\xi,\eta)$ 
as follows:
\begin{subequations}
\label{eq:config_xr_lp_zeta}
\begin{align}
    \label{eq:def_xii}
    &\xi^{(i)} = g_{\xi}(i), \quad i = 0, 1, 2, \dots, N_{\xi}, \\
    \label{eq:def_etaj}
    &\eta^{(j)} = g_{\eta}(j), \quad j = 0, 1, 2, \dots, N_{\eta}, 
\end{align}
\end{subequations}
where $g_{\xi}(y)$ and $g_{\eta}(y)$ are monotonically increasing functions that define our lattice system, i.e.,
\begin{subequations}
\label{eq:cond_gx_ge_gz}
\begin{align}
    \label{eq:cond_gx}
    &0 = g_{\xi}(0) < g_{\xi}(1) < \cdots < g_{\xi}(N_{\xi}) = \xi_{\max}, \\
    \label{eq:cond_ge}
    &0 = g_{\eta}(0) < g_{\eta}(1) < \cdots < g_{\eta}(N_{\eta}) = \pi/2, 
\end{align}
\end{subequations}
The corresponding lattice points for $x$ and $r$ are computed from \eqref{eq:oblate} as
\begin{align}
    \label{eq:def_xrij}
     x^{(i,j)} = \sinh \xi^{(i)} \cos \eta^{(j)}, \quad
     r^{(i,j)} = \cosh \xi^{(i)} \sin \eta^{(j)}.
\end{align}

The lattice points for $\zeta$, $\theta_{\zeta}$, and $\varphi_{\zeta}$ are introduced in a similar manner. However, their locations are chosen to facilitate the application of quadrature formulas for evaluating \eqref{eq:omega_cylind}--\eqref{eq:tau_cylind} and \eqref{eq:def_sigmaw}. Specifically, for $\zeta$, we first divide the interval $[0,\zeta_{\max}]$ into subintervals $[\check{\zeta}^{(k'-1)}, \check{\zeta}^{(k')}]$ for $k'=1,2,\dots,N_{\zeta}'$, where the endpoints are defined by
\begin{align}
    \label{eq:def_c_zetak}
    \check{\zeta}^{(k')} = g_{\zeta}(k'), \quad k' = 0, 1, 2, \dots, N_{\zeta}',
\end{align}
and $g_{\zeta}(y)$ is a monotonically increasing function satisfying
\begin{align}
    \label{eq:cond_gz}
    &0 = g_{\zeta}(0) < g_{\zeta}(1) < \cdots < g_{\zeta}(N_{\zeta}') = \zeta_{\max}.
\end{align}
The lattice points for $\zeta$ are then defined as the set of quadrature nodes placed within each subinterval $[\check{\zeta}^{(k'-1)}, \check{\zeta}^{(k')}]$, $k'=1,2,\dots,N_{\zeta}'$, and
 are denoted by $\zeta^{(k)}$ ($k=1,2,\dots,N_{\zeta}$).

Similarly, for the angular variables $\theta_{\zeta}$ and $\varphi_{\zeta}$, 
we define subintervals $[\check{\varphi}_{\zeta}^{(m'-1)}, \check{\varphi}_{\zeta}^{(m')}]$, $m'=1,2,\dots,N_{\varphi_{\zeta}}'$, and 
$[\check{\theta}_{\zeta}^{(l'-1)}, \check{\theta}_{\zeta}^{(l')}]$, 
$l'=1,2,\dots,N_{\theta_{\zeta}}'$.
The endpoints of these subintervals are chosen based on whether
$0 \le r^{(i,j)} < 1$ or $r^{(i,j)} \ge 1$, as the structure of the discontinuity differs between these cases. 
For $0 \leq r^{(i,j)} < 1$, we define
\begin{subequations}
\label{eq:def_phim_thetal_0r1}
\begin{align}
    \label{eq:def_phim_0r1}
    &\check{\varphi}_{\zeta}^{(m')} = g_{\varphi_{\zeta}}(m'), \quad m' = 0,1,2,\dots,N'_{\varphi_{\zeta}}, \\
    \label{eq:def_thetal_0r1}
    &\check{\theta}_{\zeta}^{(l')} = \begin{cases} g_{\theta_{\zeta}}^{-}(l'), & l' = 0,1,2,\dots,l_*^{(i,j,m')}, \\
     g_{\theta_{\zeta}}^{+}(l'), & l' = l_*^{(i,j,m')}+1,\dots,N'_{\theta_{\zeta}},
     \end{cases}
\end{align}
\end{subequations}
and $g_{\varphi_{\zeta}}(y)$, $g_{\theta_{\zeta}}^{-}(y)$, and $g_{\theta_{\zeta}}^{+}(y)$  are monotonically increasing functions satisfying
\begin{subequations}
\begin{align}
    \label{eq:cond_gp_0r1}
    0 &= g_{\varphi_{\zeta}}(0) < g_{\varphi_{\zeta}}(1) < \cdots < g_{\varphi_{\zeta}}(N'_{\varphi_{\zeta}})
       = \pi, \\
    \label{eq:cond_gt_0r1}
    0 &= g_{\theta_{\zeta}}^-(0) < g_{\theta_{\zeta}}^-(1) < \cdots < g_{\theta_{\zeta}}^-(l_*^{(i,j,m')}) \notag \\
      &= \theta_{\zeta *}^{+(i,j,m')} < g_{\theta_{\zeta}}^{+}(l_*^{(i,j,m')}+1) < \cdots
       < g_{\theta_{\zeta}}^{+}(N'_{\theta_{\zeta}}) = \pi.
\end{align}
\end{subequations}
Here, $\theta_{\zeta *}^{+(i,j,m')} = \theta_{\zeta *}^{+} (x^{(i,j)}, r^{(i,j)}, \check{\varphi}_{\zeta}^{(m')})$ (cf. \eqref{eq:def_tzastpm}).
Note that the points $\check{\theta}_{\zeta}^{(l')}$ also depend on $(i,j,m')$ through $l_*^{(i,j,m')}$ and $\theta_{\zeta *}^{+(i,j,m')}$, although this dependency is not explicitly indicated in the notation $\check{\theta}_{\zeta}^{(l')}$. A similar comment applies to $\check{\varphi}_\zeta^{(m')}$ and $\check{\theta}_\zeta^{(l')}$ in subsequent discussions and will not be repeated.
For the case of $r^{(i,j)} \geq 1$, we define
\begin{subequations}
\label{eq:def_phim_thetal_r1}
\begin{align}
    \label{eq:def_phim_r1}
    \check{\varphi}_{\zeta}^{(m')} & = \begin{cases} g_{\varphi_{\zeta}}^{-}(m'), & m' = 0,1,2,\dots,m_*^{(i,j)}, \\
    g_{\varphi_{\zeta}}^{+}(m'), & m' = m_*^{(i,j)}+1,\dots,N'_{\varphi_{\zeta}}, \end{cases} \\
    \label{eq:def_thetal_r1_mp}
    \check{\theta}_{\zeta}^{(l')} & = 
     \begin{cases} g_{\theta_{\zeta}}^{\flat}(l'), &l' = 0,1,2,\dots,l_{\dagger}^{(i,j,m)}, \\
     g_{\theta_{\zeta}}^{\natural}(l'), &l' = l_{\dagger}^{(i,j,m)}+1,\dots,l_{\dagger\dagger}^{(i,j,m)},
     \\
     g_{\theta_{\zeta}}^{\sharp}(l'), &l' = l_{\dagger\dagger}^{(i,j,m)}+1,\dots,N'_{\theta_{\zeta}},
     \end{cases}
      \qquad (m' \le m_*^{(i,j)}),\\
     \check{\theta}_{\zeta}^{(l')} & = g_{\theta_{\zeta}}(l'), \quad l' = 0,1,2,\dots,N'_{\theta_{\zeta}} \quad (m' > m_*^{(i,j)}),
\end{align}
\end{subequations}
where $g_{\varphi_{\zeta}}^{-}(y)$, $g_{\varphi_{\zeta}}^{+}(y)$, $g_{\theta_{\zeta}}^\flat(y)$, $g_{\theta_{\zeta}}^\natural(y)$, $g_{\theta_{\zeta}}^\sharp(y)$, and $g_{\theta_{\zeta}}(y)$ are monotonically increasing functions satisfying
\begin{subequations}
\label{eq:cond_gpt}
\begin{align}
    \label{eq:cond_gp_r1}
    0 & = g_{\varphi_{\zeta}}^-(0) < g_{\varphi_{\zeta}}^-(1) < \cdots < g_{\varphi_{\zeta}}^-(m_*^{(i,j)}) \notag \\
       &= \varphi_{\zeta *}^{(i,j)} 
      < g_{\varphi_{\zeta}}^+(m_*^{(i,j)}+1) < \cdots 
       < g_{\varphi_{\zeta}}^+(N'_{\varphi_{\zeta}}) = \pi, \\
    \label{eq:cond_gt_r12}
    0 & = g_{\theta_{\zeta}}^{\flat}(0) < g_{\theta_{\zeta}}^{\flat}(1) < \cdots
       < g_{\theta_{\zeta}}^{\flat}(l_{\dagger}^{(i,j,m')}) = \theta_{\zeta *}^{-(i,j,m')} < g_{\theta_{\zeta}}^{\natural}(l_{\dagger}^{(i,j,m')}+1) < \cdots 
        \nonumber \\
       & < g_{\theta_{\zeta}}^{\natural}(l_{\dagger\dagger}^{(i,j,m')}) = \theta_{\zeta *}^{+(i,j,m')} < g_{\theta_{\zeta}}^{\sharp}(l_{\dagger\dagger}^{(i,j,m')}+1) < \cdots
       < g_{\theta_{\zeta}}^{\sharp}(N'_{\theta_{\zeta}}) = \pi, \\
    \label{eq:cond_gt_r11}
    0 & = g_{\theta_{\zeta}}(0) < g_{\theta_{\zeta}}(1) < \cdots 
       < g_{\theta_{\zeta}}(N'_{\theta_{\zeta}}) = \pi,
\end{align}
\end{subequations}
where $\varphi_{\zeta *}^{(i,j)} = \varphi_{\zeta *}(r^{(i,j)})$ (cf. \eqref{eq:def_pzast}) and $\theta_{\zeta *}^{\mp(i,j,m')} = \theta_{\zeta *}^{\mp} (x^{(i,j)}, r^{(i,j)}, \check{\varphi}_{\zeta}^{(m')})$ (cf. \eqref{eq:def_tzastpm}). 
Then, the lattice points for $\theta_{\zeta}$ and those for $\varphi_{\zeta}$ are defined as the sets of quadrature nodes placed within each subinterval $[\check{\theta}_{\zeta}^{(l'-1)}, \check{\theta}_{\zeta}^{(l')}]$, $l'=1,2,\dots,N'_{\theta_{\zeta}}$, and $[\check{\varphi}_{\zeta}^{(m'-1)}, \check{\varphi}_{\zeta}^{(m')}]$, $m'=1,2,\dots,N'_{\varphi_{\zeta}}$, and are denoted by $\theta_{\zeta}^{(l)}$ ($l = 1,2,\dots,N_{\theta_{\zeta}}$) and $\varphi_{\zeta}^{(m)}$ ($m=1,2,\dots,N_{\varphi_{\zeta}}$), respectively.

We introduce the notation for the discretized $\phi_C$ as
\begin{align}
    \label{eq:descri_VDF_macro_sigma}
    & \phi_{C,ijklm} = \phi_C( x^{(i,j)}, r^{(i,j)}, \zeta^{(k)}, \theta_{\zeta}^{(l)},\varphi_{\zeta}^{(m)}).
\end{align}
Similarly, the values of $\omega, \, u_x, \, u_r, \, \tau$, and $\sigma_{\mathrm{w}}$ at the lattice points are expressed as
\begin{align}
    & h_{ij} = h(x^{(i,j)}, r^{(i,j)}) \quad (h = \omega, \, u_x, \, u_r, \, \tau),
    \quad
    \sigma_{\mathrm{w},j} = \sigma_{\mathrm{w}}(r^{(0,j)}).
\end{align}
The discretized value $\phi_{C,ijklm}$ is obtained as the limit of the sequence $\{\phi_{C,ijklm}^{(n)}\}$, where $n=0,1,2,\ldots$, produced by successively applying the following scheme, starting from a suitably chosen initial value $\phi_{C,ijklm}^{(0)}$:
\begin{align}
    \label{eq:update_rule}
    \phi_{C,ijklm}^{(n+1)} & = 
    \displaystyle \exp( -\frac{s_{\mathrm{w},ijlm}}{\kappa \zeta^{(k)}} ) \sigma_{\mathrm{w}}^{(n)}(r_{\mathrm{w},ijlm}) \mathbbm{1}_{\Omega,ijlm}
    \nonumber
    \\[0.3cm] 
    & + \frac{1}{\kappa \zeta^{(k)}} \int_{0}^{s_{*,ijlm}} \exp( -\frac{s}{\kappa \zeta^{(k)}} )
      G^{(n)}(\tilde{x}_{ijlm}(s), \tilde{r}_{ijlm}(s), \zeta^{(k)}, \theta_{\zeta}^{(l)},
              \tilde{\varphi}_{\zeta,ijlm}(s)) \mathrm{d}s,
\end{align}
where 
\begin{align}
    \label{eq:def_swijl}
    &s_{\mathrm{w},ijlm} = s_{\mathrm{w}}(x^{(i,j)}, \theta_{\zeta}^{(l)}), \quad
    r_{\mathrm{w},ijlm} = r_{\mathrm{w}}(x^{(i,j)}, r^{(i,j)}, \theta_{\zeta}^{(l)}, \varphi_{\zeta}^{(m)}), \\
    & (\mathbbm{1}_{\Omega,ijlm},s_{*,ijlm}) = \begin{cases}
        (1,s_{\mathrm{w},ijlm}), \quad (x^{(i,j)},r^{(i,j)},\theta_\zeta^{(l)},\varphi_\zeta^{(m)}) \in \Omega, \\
        (0,s_{\infty,ijlm}), \quad (x^{(i,j)},r^{(i,j)},\theta_\zeta^{(l)},\varphi_\zeta^{(m)}) \notin \Omega,
    \end{cases}
    \\
    & \tilde{x}_{ijlm}(s) = \tilde{x}(s; x^{(i,j)},\theta_{\zeta}^{(l)}), \quad
    \tilde{r}_{ijlm}(s) = \tilde{r}(s; r^{(i,j)}, \theta_{\zeta}^{(l)}, \varphi_{\zeta}^{(m)}),
    \\
    & \tilde{\varphi}_{\zeta, ijlm}(s) = \tilde{\varphi}_{\zeta}(s; r^{(i,j)},\theta_{\zeta}^{(l)},\varphi_{\zeta}^{(m)}).
\end{align}
In this scheme, $s_{\infty,ijlm}$ is a sufficiently large positive number, ensuring numerical convergence of the integral over the infinite range.
Note that $G^{(n)}$ in \eqref{eq:update_rule} is defined by \eqref{eq:def_G}, with the following substitutions:
$\omega(\tilde{x},\tilde{r})=\omega^{(n)}(\tilde{x},\tilde{r})$, $u_x(\tilde{x},\tilde{r})=u_x^{(n)}(\tilde{x},\tilde{r})$, $u_r(\tilde{x},\tilde{r})=u_r^{(n)}(\tilde{x},\tilde{r})$, and $\tau(\tilde{x},\tilde{r})=\tau^{(n)}(\tilde{x},\tilde{r})$, where $\tilde{x}=\tilde{x}_{ijlm}(s)$ and $\tilde{r}=\tilde{r}_{ijlm}(s)$.

To evaluate the integral with respect to $s$ in \eqref{eq:update_rule}, the interval $[0,s_{*,ijlm}]$ is divided into subintervals.
The Gauss--Legendre four-point formula is then applied to each subinterval, and the results are summed.
In these processes, the macroscopic quantities $\omega^{(n)}$, $u_x^{(n)}$, $u_r^{(n)}$, and $\tau^{(n)}$ are interpolated from their values at the lattice points using the standard second-order Lagrange formula (performed first along the $\xi$ variable and then along the $\eta$ variable successively).
The value of $\sigma_{\mathrm{w}}^{(n)}(r_{\mathrm{w},ijlm})$ in \eqref{eq:update_rule} is  interpolated from the values of $\sigma_{\mathrm{w},j}^{(n)}$ at the $n$-th iteration using the second-order Lagrange formula.

Once $\phi_C$ is computed at the $(n+1)$-th step, the quantities $\omega_{ij}^{(n+1)}$, $u_{x,ij}^{(n+1)}$, $u_{r,ij}^{(n+1)}$, $\tau_{ij}^{(n+1)}$, and $\sigma_{\mathrm{w},j}^{(n+1)}$ are calculated from $\phi_{C,ijklm}^{(n+1)}$ using Eqs.~\eqref{eq:omega_cylind}--\eqref{eq:tau_cylind} and \eqref{eq:def_sigmaw}. These quantities are obtained by applying the Gauss--Legendre four-point formula.

\subsection{Asymptotic behavior in the far field}

In this subsection, we discuss the asymptotic behavior of the flow in the far-field, which is used to enhance the accuracy of numerical computations. Suppose the deviation $\phi - 2 \zeta_1 u_{\infty}$ decays in proportion to the reciprocal of $\hat{r} = \hat{r}(x,r) = \sqrt{x^2+r^2}$, the distance from the origin. If this is the case, 
the effective (or local) Knudsen number is small if $\hat{r}(x,r) > \hat{r}_A$, where $\hat{r}_A$ satisfies $0 <\kappa/\widetilde{r}_A \ll 1$.
This implies that the asymptotic theory of the BGK equation (or the Boltzmann equation) for small Knudsen numbers \citep{Sone02,Sone07} can be applied to derive asymptotic expressions for the flow in the far field. 
As a result, the density, flow velocity, temperature, and pressure are expressed in terms of oblate spheroid coordinates as follows (see Sec.~S1 of Supplementary Materials for details of derivation):
\begin{subequations}
\label{eq:far_asy}
\begin{align}
    \label{eq:omega_far}
    &\frac{\omega}{u_{\infty}} = -\frac{(2 c_1 \kappa + c_3) \cos\eta}{\sinh^2 \xi + \cos^2 \eta}, \\
    \label{eq:ux_far}
    &\frac{u_x}{u_{\infty}}
    = 1 - 2 c_2 \text{\,arccot} (\sinh \xi) + \frac{\sinh \xi}{\sinh^2\xi + \cos^2\eta} 
      \left[ -c_1 (1+\cos^2 \eta) + 2c_2 \right], \\
    \label{eq:ur_far}
    &\frac{u_r}{u_{\infty}}
    = \frac{\cosh \xi \sin 2\eta}{\sinh^2 \xi + \cos^2 \eta} 
      \left(\frac{c_1-c_2}{\cosh^2\xi} - \frac{c_1}{2} \right), \\
    \label{eq:tau_far}
    &\frac{\tau}{u_{\infty}} = \frac{c_3 \cos \eta}{\sinh^2 \xi + \cos^2 \eta},
    \\
    &\frac{P}{u_{\infty}} = -\frac{2 c_1 \gamma_1 \kappa \cos\eta}{\sinh^2 \xi + \cos^2 \eta}. 
    \label{eq:p_far}
\end{align}
\end{subequations}
Here, $\gamma_1$ represents the dimensionless viscosity, which relates the viscosity at the reference state $\mu_0$ through the expression $\mu_0 = \gamma_1 \kappa p_0L/(2RT_0)^{1/2}$.
For the present BGK model, $\gamma_1=1$ \citep[e.g., ][]{Sone07}.
The constants $c_1$, $c_2$, and $c_3$ are arbitrary parameters determined by matching the asymptotic expressions with the numerical solution in a region far from the disk. It is important to note that these constants $c_i$ depend on $\kappa$ (the Knudsen number), as the matching process reflects the influence of $\kappa$ on the solution. 

Suppose the constants $c_i$ are known. The asymptotic forms \eqref{eq:far_asy} are then applied to evaluate the integrand in Eq.~\eqref{eq:inteq_unfied} when the argument $(\widetilde{x}(s),\widetilde{r}(s))$ lies outside the computational domain (i.e., $\xi > \xi_{\max})$. It is worth noting that, in our integral equation, only the macroscopic quantities are required to evaluate the integrand, meaning that the asymptotic forms of the VDF are not necessary.
Further details of the numerical analysis, as well as the matching process, are provided in Appendix~\ref{app:matching}. 

In the far field, the terms in Eq.~\eqref{eq:far_asy} involving the constant $c_1$ can be interpreted as a Stokeslet. This becomes evident when \eqref{eq:ux_far} and \eqref{eq:ur_far} are expanded in terms of $\chi=e^{-\xi} \sim (2 \hat{r})^{-1}$. Based on this consideration, the following relation can be derived:
\begin{align} \label{eq:relation_hd_c1}
    h_D = 8 \pi \gamma_1 \kappa c_1.
\end{align}
This identity serves as a measure of the accuracy of the present computations.

\section{\label{sec:freemolecular}Case of a free molecular gas}
Before presenting our numerical results, we consider the case of a free molecular gas, corresponding to the limit $\kappa \to \infty$.
In this case, the solution is given by
\begin{align}
\label{eq:phiC_fmg}
    \phi_C(x,r,\zeta,\theta_{\zeta},\varphi_{\zeta}) = \begin{cases}
    \sqrt{\pi} u_{\infty}, & (x,r,\zeta,\theta_{\zeta},\varphi_{\zeta}) \in \Omega, \\
    2\zeta u_{\infty} \cos \theta_{\zeta}, & (x,r,\zeta,\theta_{\zeta},\varphi_{\zeta}) \notin \Omega.
    \end{cases}
\end{align}
The macroscopic quantities are calculated using this distribution function.
In particular, the normal stress component $P_{xx}$ on the disk ($x=0_+$) is obtained as
\begin{align}
\label{eq:pxx_fmg}
    \frac{P_{xx}(x=0_+,r)}{u_{\infty}} = -\frac{\pi+4}{2\sqrt{\pi}} \qquad (0\leq r < 1).
\end{align}
Thus, the normal stress is uniform with respect to $r$ on the disk. Substituting this into \eqref{eq:def_h_D}, the force acting on the disk in the free-molecular limit is given by
\begin{align}
\label{eq:h_D_fmg}
    h_D(\infty) = \sqrt{\pi}(\pi+4).
\end{align}
The expression coincides with that obtained in \citet{Dahneke_1973} under the present flow conditions (i.e., fully diffusive reflection and no temperature difference between the disk and the gas at infinity).

\section{\label{sec:numerical results}Numerical results}

We have carried out numerical computations as described above, varying $\kappa$ from 0.02 to 10. This section presents the corresponding numerical results. Supporting data regarding the accuracy of the computations are provided in Appendix~\ref{app:accuracy}.

\begin{figure*}
\begin{center}
\subfigure[$r=0$]{\includegraphics[scale=0.65]{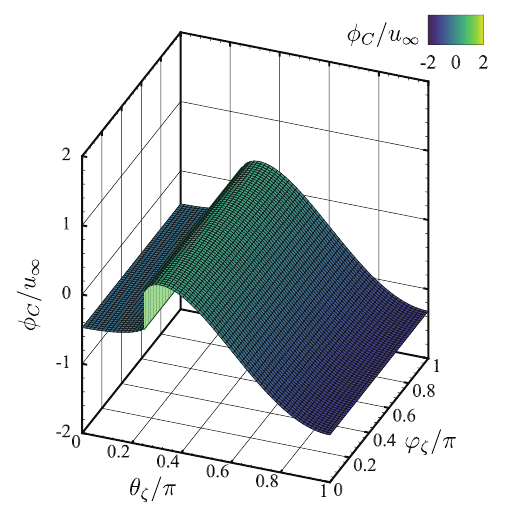}}
\qquad
\subfigure[$r=0.5$]{\includegraphics[scale=0.65]{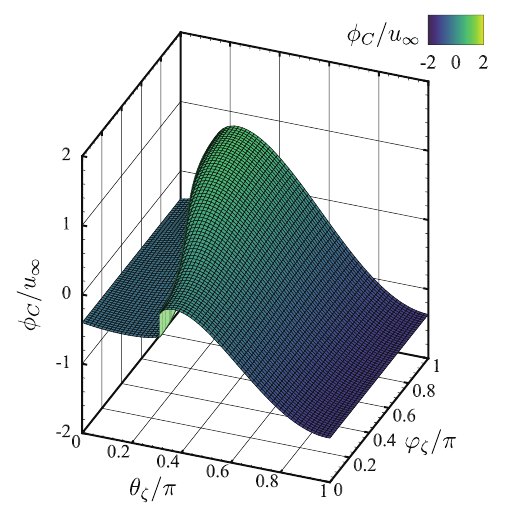}}
\subfigure[$r=0.75$]{\includegraphics[scale=0.65]{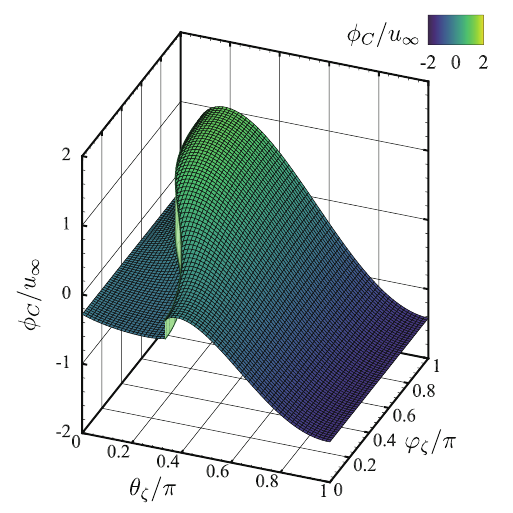}}
\qquad
\subfigure[$r=1$]{\includegraphics[scale=0.65]{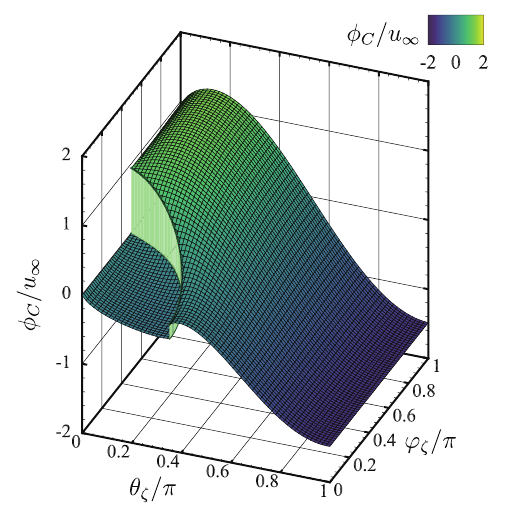}}
\subfigure[$r=1.5$]{\includegraphics[scale=0.65]{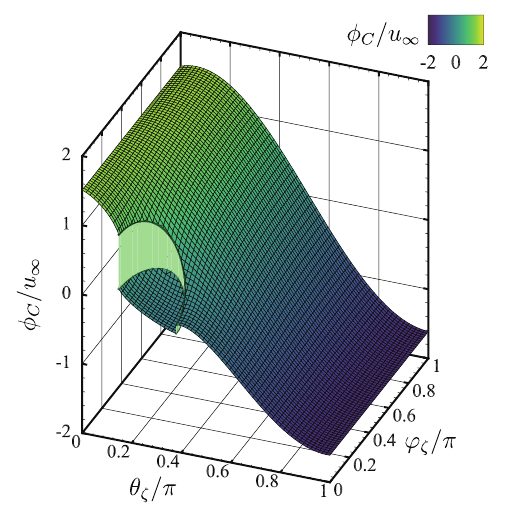}}
\qquad
\subfigure[$r=2$]{\includegraphics[scale=0.65]{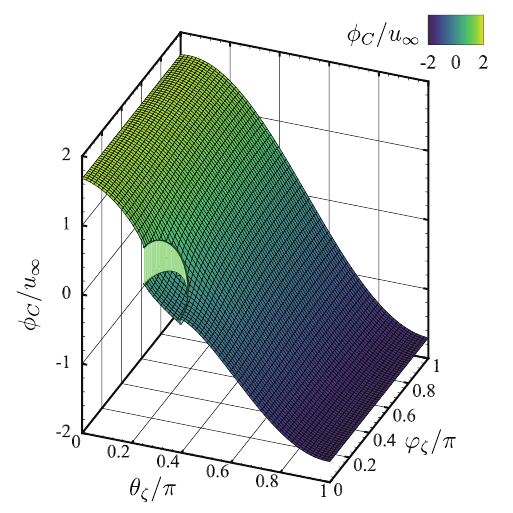}}
\caption{$\phi_C(x,r,\theta_\zeta,\varphi_\zeta,\zeta)$ as a function of $\theta_\zeta$ and $\varphi_\zeta$ for $x=1$ and $\zeta=1$ and for various $r$ in the case of $\kappa=1$.
(a) $r=0$, (b) $r=0.5$, (c) $r=0.75$, (d) $r=1$, (e) $r=1.5$, (f) $r=2$.}
\label{fig:vdf_k1}
\end{center}
\end{figure*}
\begin{figure*}
\begin{center}
\subfigure[ $r=0$]{\includegraphics[scale=0.65]{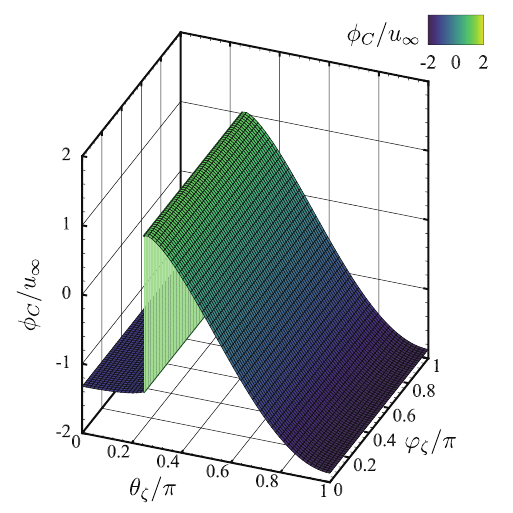}}
\qquad
\subfigure[$r=0.5$]{\includegraphics[scale=0.65]{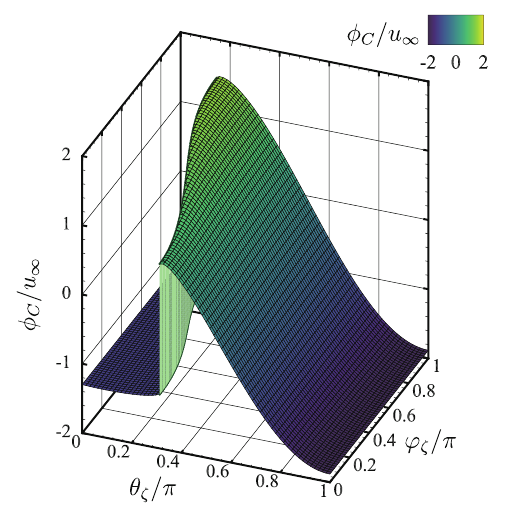}}
\subfigure[$r=0.75$]{\includegraphics[scale=0.65]{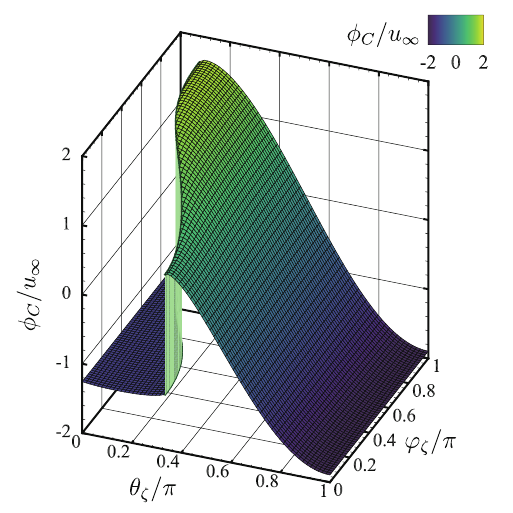}}
\qquad
\subfigure[$r=1$]{\includegraphics[scale=0.65]{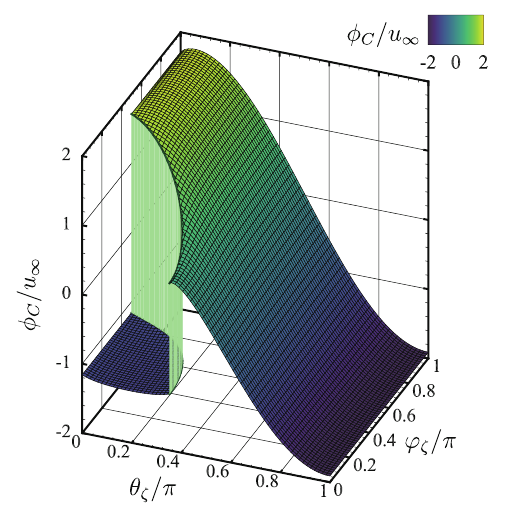}}
\subfigure[$r=1.5$]{\includegraphics[scale=0.65]{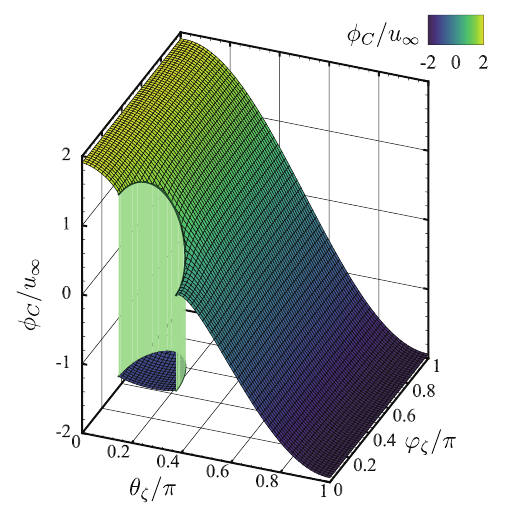}}
\qquad
\subfigure[$r=2$]{\includegraphics[scale=0.65]{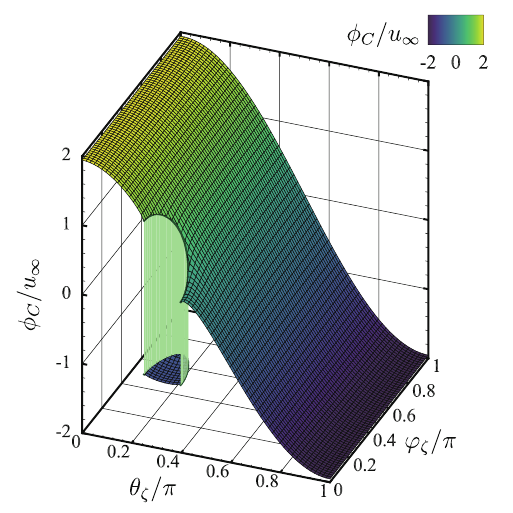}}
\caption{$\phi_C(x,r,\theta_\zeta,\varphi_\zeta,\zeta)$ as a function of $\theta_\zeta$ and $\varphi_\zeta$ for $x=1$ and $\zeta=1$ and for various $r$ in the case of $\kappa=5$.
(a) $r=0$, (b) $r=0.5$, (c) $r=0.75$, (d) $r=1$, (e) $r=1.5$, (f) $r=2$.}
\label{fig:vdf_k5}
\end{center}
\end{figure*}

\subsection{\label{subsec:VDF}Velocity distribution function}

We begin by examining the behavior of the VDF in Figs.~\ref{fig:vdf_k1} and \ref{fig:vdf_k5}.
Figure~\ref{fig:vdf_k1} shows $\phi_C/u_{\infty}$ as a function of $\theta_{\zeta}$ and $\varphi_{\zeta}$ at $x=1$ and $\zeta=1$ for various $r$ ($0 \le r \le 2$) in the case of $\kappa=1$, while Fig.~\ref{fig:vdf_k5} shows the corresponding data for $\kappa=5$. The location $x = 1$ is selected for illustrative purposes; note that both the position and the shape of the discontinuity vary with $x$.
These figures clearly demonstrate the discontinuities in the VDF.
For $r<1$ [(a,b,c)], the VDF exhibits a jump at $\theta_{\zeta} = \theta_{\zeta *}^{+}$ for each value of $\varphi_{\zeta} \in [0,\pi]$. In contrast, for $r>1$ [(e,f)],
two jumps occur at $\theta_{\zeta} = \theta_{\zeta *}^{+}$ and $\theta_{\zeta *}^{-}$ for each  $\varphi_{\zeta} (< \varphi_{\zeta *})$ (cf. Fig.~\ref{fig:discontinuity}).
For the case of $r=1$ [(d)], a jump is observed at $\theta_{\zeta} = \theta_{\zeta *}^{+}$ for each $\varphi_{\zeta} < \pi/2$.

The discontinuities diminish with increasing distance from the disk edge due to molecular collisions. As a result, the magnitudes of the jumps decrease for larger values of $r$ 
(specially when $r \ge 1$) at a fixed $\kappa$. As $\kappa$ increases, the mean free path becomes larger. Consequently, for the same value of $r$, the effective distance from the edge is reduced, making the discontinuity more pronounced at higher $\kappa$.
Additionally, the area of $\Omega_2$ (see Eq.~\eqref{eq:def_Omega_2}) shrinks as $r$ increases (see the panels (d,e,f)).
In summary, our numerical analysis effectively captures the characteristic behavior of the VDF.


\begin{figure*}
\begin{center}
\subfigure[]{\includegraphics[scale=0.77]{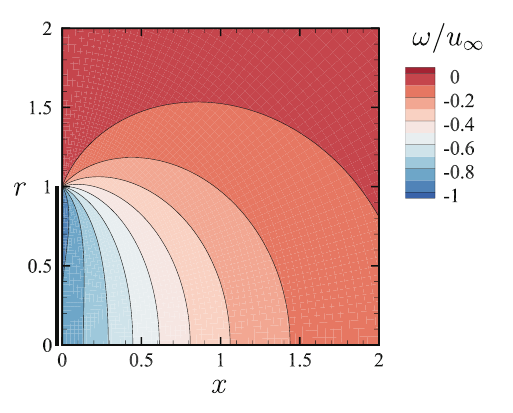}}
\subfigure[]{\includegraphics[scale=0.77]{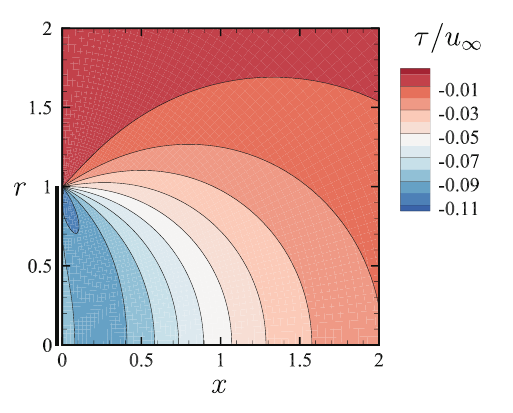}}
\subfigure[]{\includegraphics[scale=0.77]{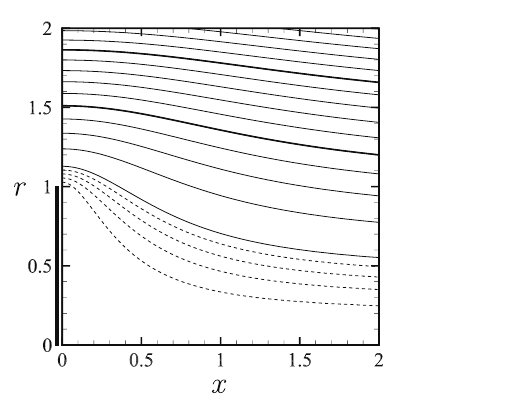}}
\subfigure[]{\includegraphics[scale=0.77]{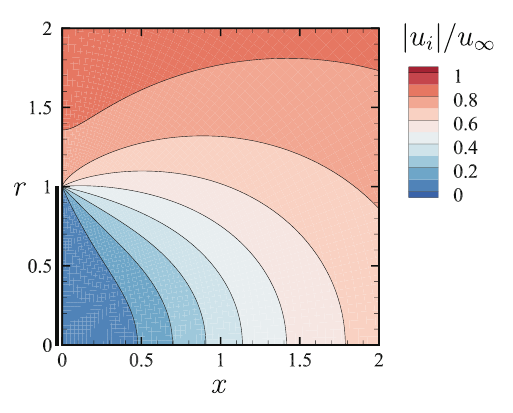}}
\caption{The behavior of macroscopic quantities around the disk in the case of $\kappa=1$. 
(a) Isolines of $\omega/u_\infty$, 
(b) isolines of $\tau/u_\infty$, 
(c) streamlines of $(u_x,u_r)/u_\infty$,
(d) isolines of $|u_i|/u_\infty = (u_x^2+u_r^2)^{1/2}/u_\infty$. 
In (c), the streamlines of $(u_x,u_r)$ are shown as isolines of the stream function $\psi$ defined in the main text.
The values of $\psi$ are $\psi=0.02 m$ ($m=1,2,3,4$) for the broken curves and $\psi=0.1 m$ ($m=1,2,\dots$) for the solid curves, where the thick solid curves are used for $\psi=0.5$ and $1$. Note that $\psi=0$ on the $x$ axis.}
\label{fig:macro_k1}
\end{center}
\end{figure*}

\begin{figure*}
\begin{center}
\subfigure[]{\includegraphics[scale=0.77]{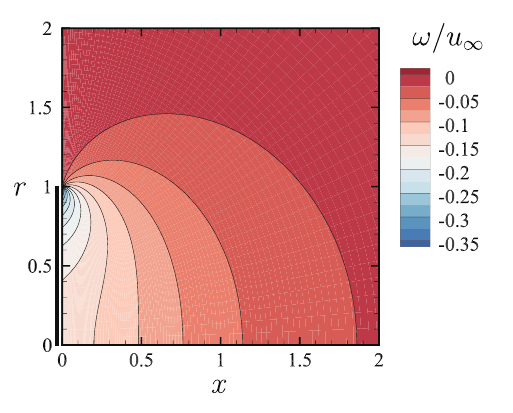}}
\subfigure[]{\includegraphics[scale=0.77]{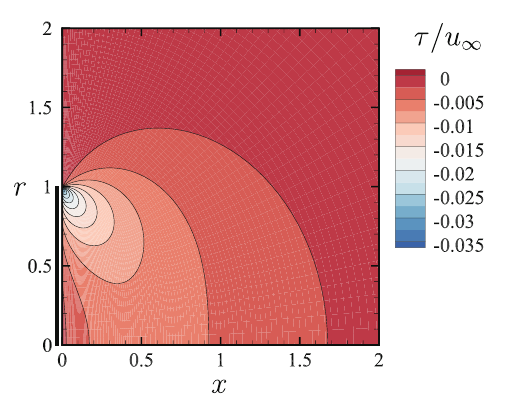}}
\subfigure[]{\includegraphics[scale=0.77]{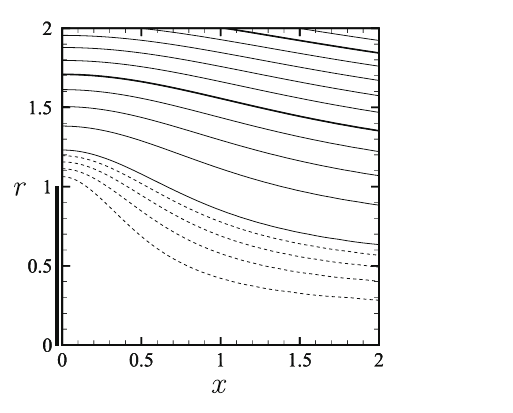}}
\subfigure[]{\includegraphics[scale=0.77]{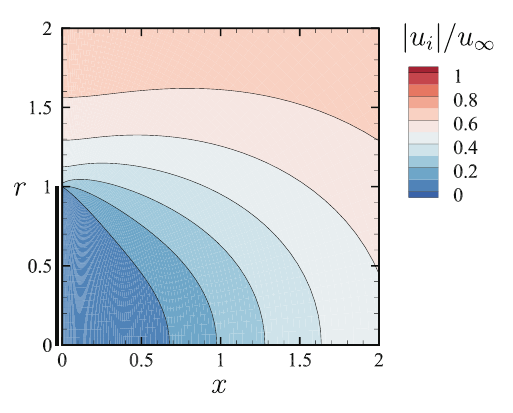}}
\caption{The behavior of macroscopic quantities around the disk in the case of $\kappa=0.1$. 
(a) Isolines of $\omega/u_\infty$, 
(b) isolines of $\tau/u_\infty$, 
(c) streamlines of $(u_x,u_r)/u_\infty$, 
(d) isolines of $|u_i|/u_\infty=(u_x^2+u_r^2)^{1/2}/u_\infty$. 
See the caption of Fig.~\ref{fig:macro_k1}.}
\label{fig:macro_k02}
\end{center}
\end{figure*}

\subsection{\label{subsec:macroscopic_quantities}Macroscopic quantities}

We now present the results for the macroscopic quantities.
Owing to the symmetry with respect to $x=0$, we will show the behavior of the macroscopic quantities for $x>0$ with the corresponding values for $x<0$ obtained using the following relations:
\begin{align}
\label{eq:macro_sym}
    h(x,r) =
\begin{cases}
    h(-x,r), &(h = u_x, P_{xr}), \\
   -h(-x,r), &(h = \omega, u_r, \tau, P, P_{xx}, P_{rr}, P_{\theta\theta}).
\end{cases}
\end{align}


Figures~\ref{fig:macro_k1} and \ref{fig:macro_k02} illustrate typical flow patterns around the disk. More precisely. Figs.~\ref{fig:macro_k1} and \ref{fig:macro_k02} show (a) the isolines of the density $\omega$, (b) those of the temperature $\tau$, and (c,d) the streamlines and magnitude of the flow velocity $(u_x,u_r)$ for $\kappa=1$ and 0.1, respectively. 
In panel (c), the isolines of the stream function $\psi$ are shown as streamlines and the flow direction is from left to right. Here, $\psi$ is defined by $r^{-1} \partial \psi/\partial r = u_x/u_\infty$ and $r^{-1} \partial \psi/\partial x= -u_r/u_\infty$. Note that the pressure is obtained through $P=\omega + \tau$ and thus omitted.
The isolines of density and temperature are more concentrated near the tip of the disk, indicating abrupt changes in macroscopic quantities in this region. At the tip, the values of the macroscopic variables are not uniquely determined and depend on the angle at which the tip is approached.
The (deviational) density $\omega/u_\infty$ and temperature $\tau/u_\infty$ take negative (or positive) values on the right (or left) side of the disk. These spatial variations become more pronounced as $\kappa$ increases. This nonuniform temperature distribution around a body in a rarefied flow is known as the thermal polarization.
The streamlines exhibit a pronounced bend near the tip of the disk, resulting in a substantial velocity gradient in that region.
As seen from the comparison of panel (d) between Figs.~\ref{fig:macro_k1} and \ref{fig:macro_k02}, the flow speed is lower for $\kappa=0.1$ than for $\kappa=1$ near the disk.

\begin{figure*}
\begin{center}
\subfigure[]{\includegraphics[scale=0.46]{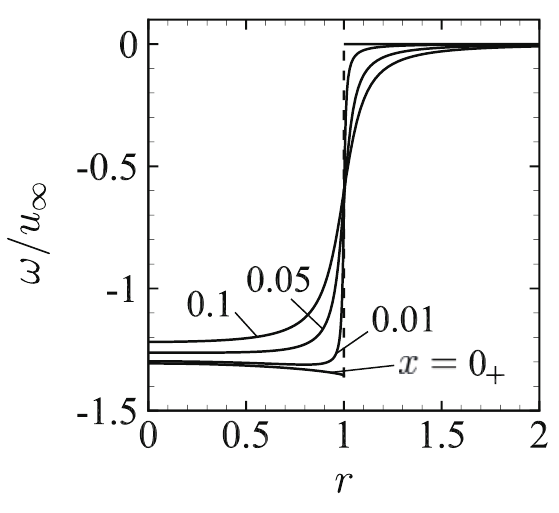}}
\subfigure[]{\includegraphics[scale=0.46]{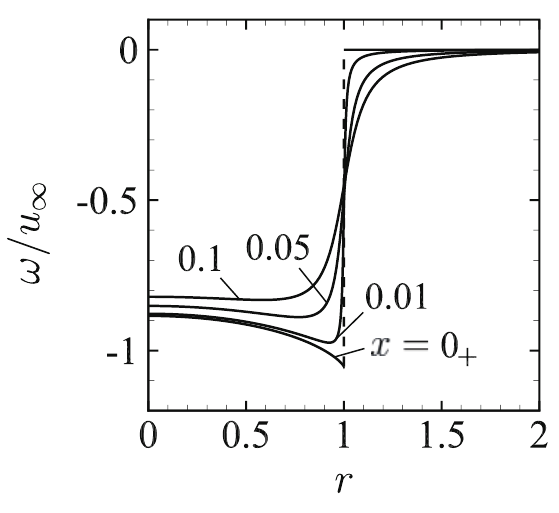}}
\subfigure[]{\includegraphics[scale=0.46]{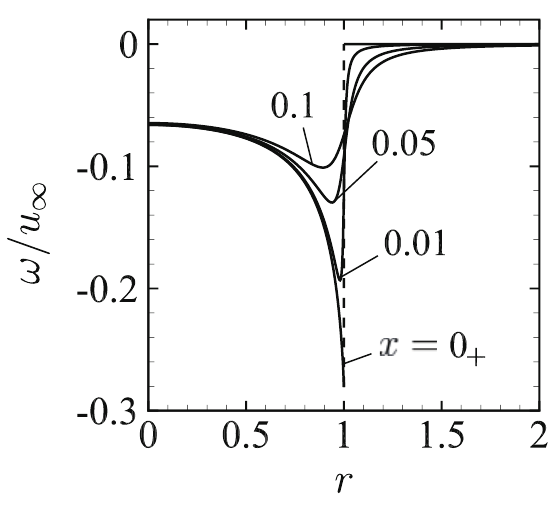}}
\subfigure[]{\includegraphics[scale=0.46]{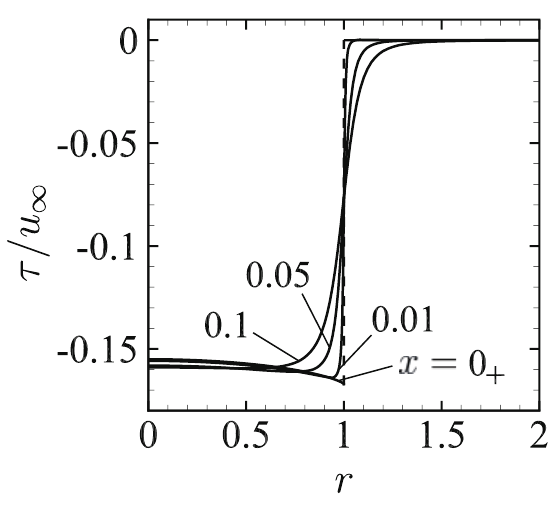}}
\subfigure[]{\includegraphics[scale=0.46]{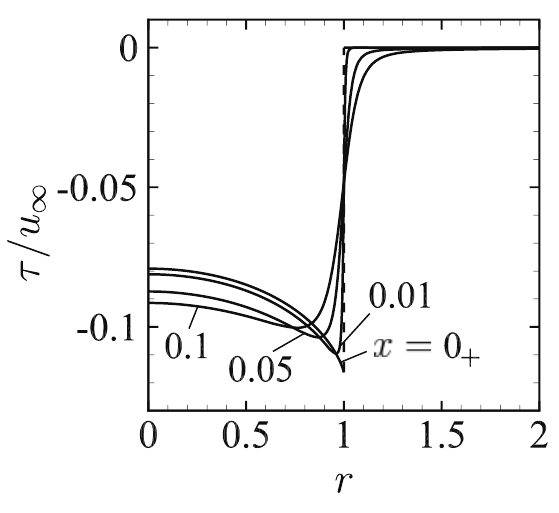}}
\subfigure[]{\includegraphics[scale=0.46]{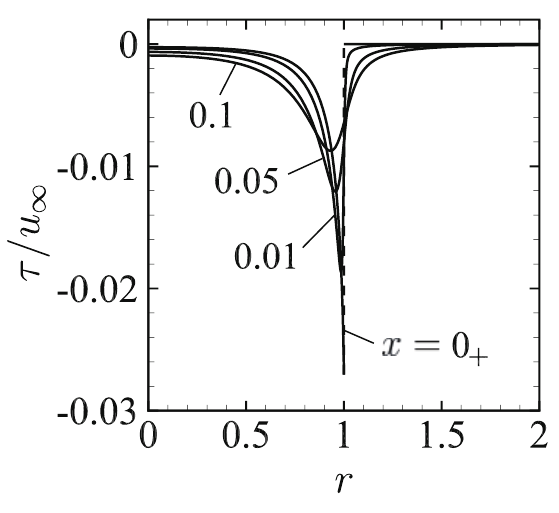}}
\subfigure[]{\includegraphics[scale=0.46]{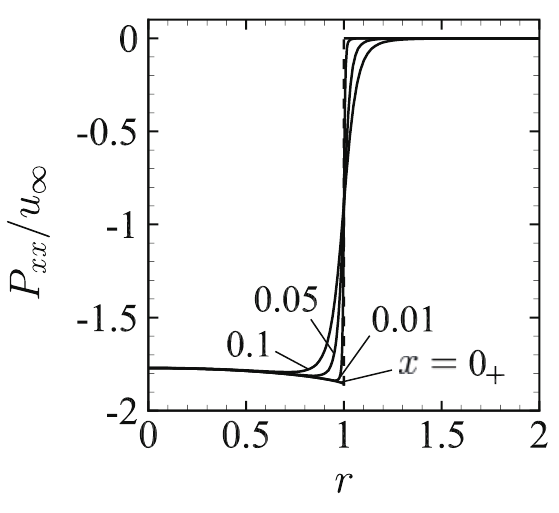}}
\subfigure[]{\includegraphics[scale=0.46]{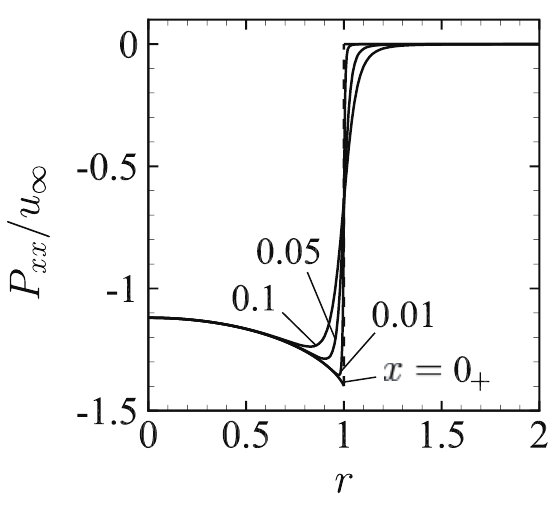}}
\subfigure[]{\includegraphics[scale=0.46]{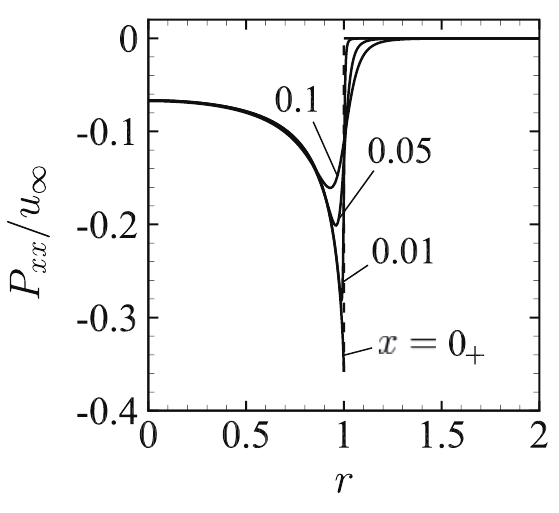}}
\caption{Profiles of $\omega/u_\infty$, $\tau/u_\infty$, and $P_{xx}/u_\infty$ along the lines $x=0_+$, 0.01, 0.05, and 0.1.
(a,d,g) $\kappa=5$, (b,e,h) $\kappa=1$, (c,f,i) $\kappa=0.05$.
The curve is discontinuous at $r=1$ along $x=0_+$, which is indicated by the dashed line.}
\label{fig:macro_near}
\end{center}
\end{figure*}

To provide a closer view of the fluid behavior near the disk,
Fig.~\ref{fig:macro_near} shows the profiles of $\omega/u_\infty$, $\tau/u_\infty$, and $P_{xx}/u_\infty$ along the lines $x=0_+$, 0.01, and 0.05 for $\kappa=5$, 1, and 0.05.
Note that the curves along $x=0_+$ exhibit discontinuities at $r=1$.
For $\kappa=5$, these quantities are nearly uniform along the disk (they are exactly uniform when $\kappa=\infty$; see Eqs.~\eqref{eq:phiC_fmg} and \eqref{eq:pxx_fmg}). As $\kappa$ decreases, the values near the central part of the disk increase (the values are negative), resulting in a more pronounced variation along the disk. For $\kappa=0.05$, a peak-like profile develops near the edge for each quantity. At this value of $\kappa$, the temperature distribution is almost uniform in the gas except in the vicinity of the disk edge.
We shall discuss the peak-like behavior near the edge below in Sec.~\ref{sec:discussions}.

The temperature is lower ($\tau <0$) on the right-hand (downstream) side of the disk and higher ($\tau >0$) on the left-hand (upstream) side. This phenomenon exemplifies thermal polarization \citep{Aoki-Sone-PHF-1987,Bakanov+Vysotskij+Derjaguin+Roldughin_JNE_1983,Takata-Sone-Aoki_PHF93}. The qualitative pattern of temperature variation is consistent with that observed for spherical bodies \citep{Takata-Sone-Aoki_PHF93}. The mechanism of thermal polarization can easily be understood in the case of a free molecular flow \citep[see also][]{Takata-Sone-Aoki_PHF93}.
Consider a disk at rest in a quiescent gas. The molecules reflected from the disk form a half-Maxwellian distribution according to the diffuse reflection condition, while the molecules incident on the disk follow a stationary Maxwellian with the density and temperature prescribed at infinity.
Now, imagine that the disk is moving through the gas perpendicularly to the disk. In the reference frame moving with the disk, the upstream Maxwellian is shifted in the molecular velocity space by an amount corresponding to the uniform flow speed. This shift causes the velocity distribution function on the upstream side of the disk to become broader, while on the downstream side, it becomes narrower. Consequently, the effective gas temperature increases on the upstream side and decreases on the downstream side of the disk. We will discuss the thermal polarization in the near-continuum regime in Sec.~\ref{sec:discussions}.



%

\begin{figure}
    \centering
    \subfigure[]{\includegraphics[width=0.45\linewidth]{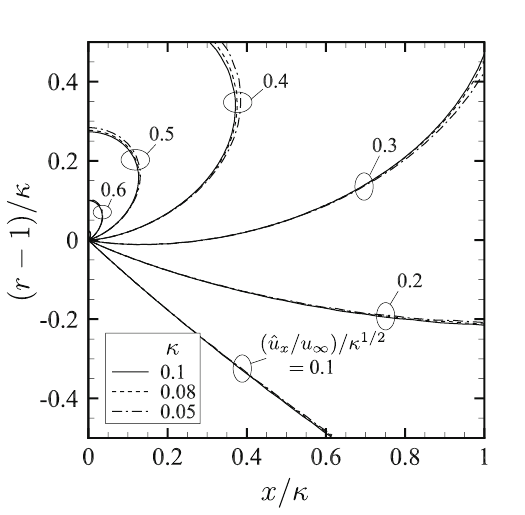}}
    \subfigure[]{\includegraphics[width=0.45\linewidth]{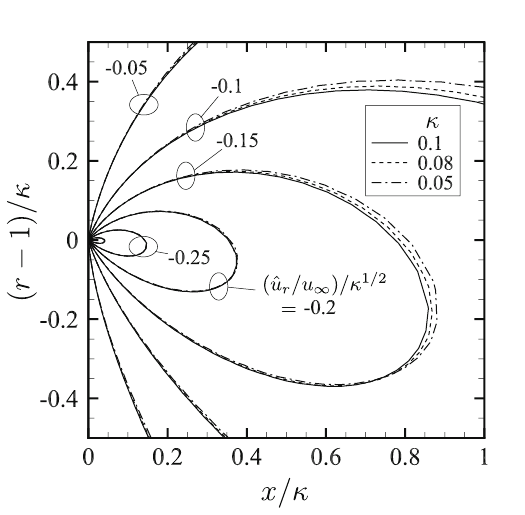}}
    \caption{Isolines of the scaled flow velocity $(\hat{u}_x/\kappa^{1/2},\hat{u}_r/\kappa^{1/2})$ superimposed for various values of  $\kappa$. Here, $\hat{u}_x = u_x - u_x^{\mathrm{St}}$ and $\hat{u}_r = u_r - u_r^{\mathrm{St}}$. (a) $\hat{u}_x/\kappa^{1/2}$, (b) $\hat{u}_r/\kappa^{1/2}$. The spatial variables $(x,r)$ are stretched around the tip by the factor of $\kappa$.}
    \label{fig:scaled_velocity}
\end{figure}
\begin{figure}
    \centering
    \subfigure[]{\includegraphics[width=0.45\linewidth]{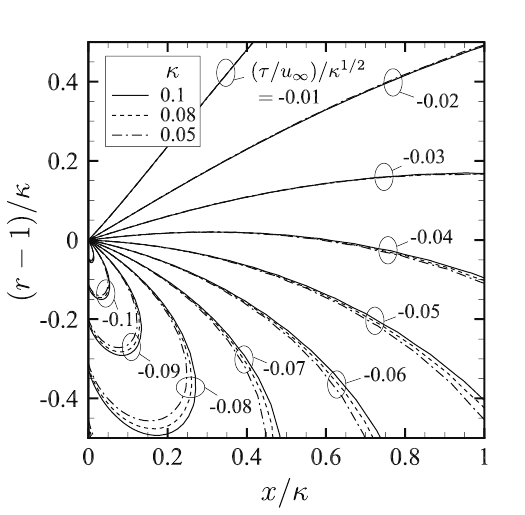}}
    \subfigure[]{\includegraphics[width=0.45\linewidth]{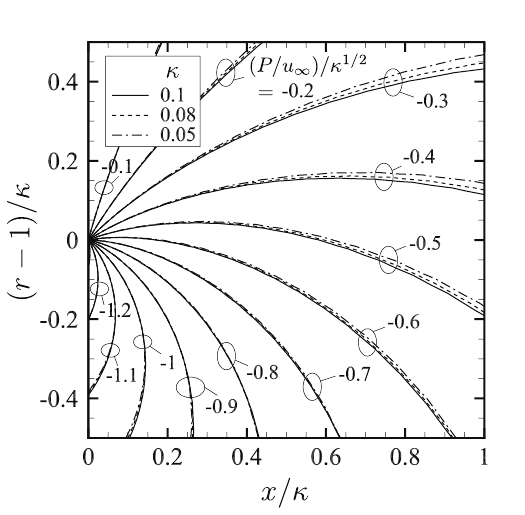}}
    \caption{Isolines of the scaled temperature $\tau/\kappa^{1/2}$ and those for scaled pressure $P/\kappa^{1/2}$ superimposed for various values of $\kappa$. See the caption of Fig.~\ref{fig:scaled_velocity}.
    }
    \label{fig:scale_tau_mod_ol_ver2}
\end{figure}

\begin{figure}
    \centering
    \subfigure[]{\includegraphics[width=0.45\linewidth]{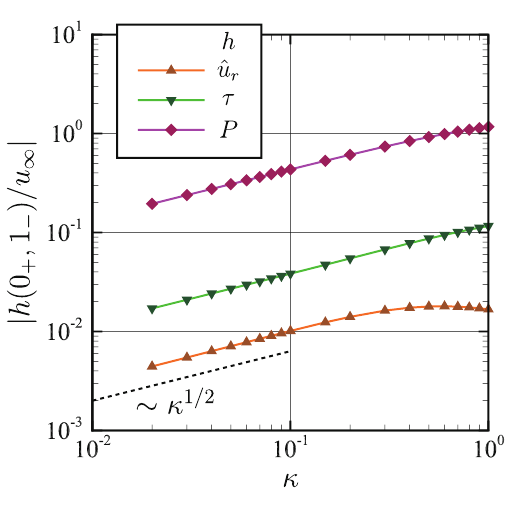}}
    \subfigure[]{\includegraphics[width=0.45\linewidth]{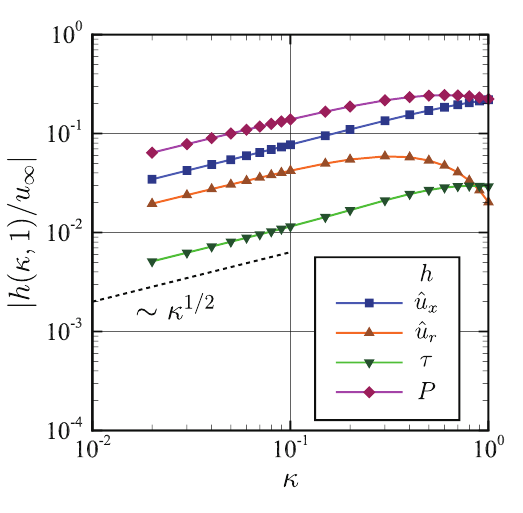}}
    \subfigure[]{\includegraphics[width=0.45\linewidth]{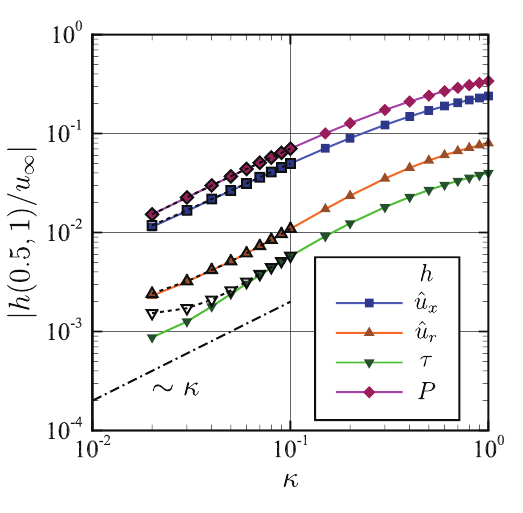}}
    \subfigure[]{\includegraphics[width=0.45\linewidth]{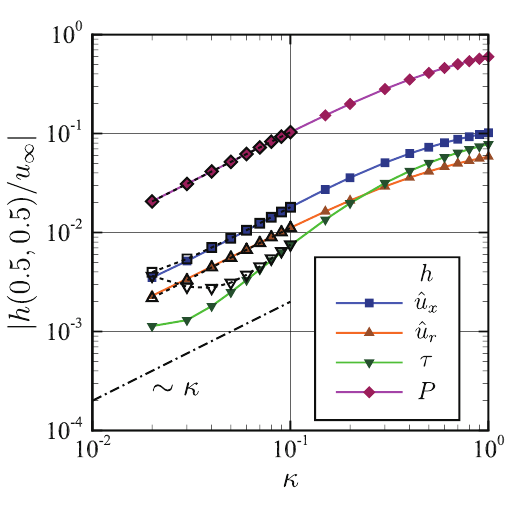}}
    \caption{Log-log plot of $h(x,r)$ ($h=\hat{u}_x$, $\hat{u}_r$, $\tau$, and $P$) versus $\kappa$ for various $(x,r)$. (a) $h(0_+,1_-) = \lim_{\epsilon \downarrow 0}h(0_+,1-\epsilon)/u_\infty$, (b) $h(\kappa,1)/u_\infty$, (c) $h(0.5,1)/u_\infty$, (d) $h(0.5,0.5)/u_\infty$. The result for $\hat{u}_x$ is not shown in panel (a) because $\hat{u}_x=0$ holds identically on the disk surface. In (c) and (d), results with coarser grids are overplotted for $\kappa \le 0.1$ using open symbols.}
    \label{fig:macro_decay}
\end{figure}

\section{\label{sec:discussions}Discussions}

\subsection{\label{subsec:edge_layer}Edge layer}
If the present flow problem is considered based on the Stokes equation with no-slip boundary conditions, the entire flow field is given by Eqs.~\eqref{eq:ux_far}, \eqref{eq:ur_far}, and \eqref{eq:p_far} with $c_1=2/\pi$ and $c_2=1/\pi$ (cf. Eq.~(S1.44) of Supplementary Materials). 
Note that the pressure as well as the derivative of the flow velocity diverge at the edge for the Stokes flow, while they remain finite in the present computation.
Let us denote them by $u_x^{\mathrm{St}}$, $u_r^{\mathrm{St}}$, and $P^{\mathrm{St}}$, respectively.
To study the behavior of the flow field near the edge when $\kappa$ is small, Fig.~\ref{fig:scaled_velocity} presents the deviation of the flow velocity $(u_x,u_r)$ from the Stokes value $(u_x^{\mathrm{St}},u_r^{\mathrm{St}})$, i.e.,
$(\hat{u}_x,\hat{u}_r)\equiv(u_r - u_r^{\mathrm{St}},u_x - u_x^{\mathrm{St}})$ for $\kappa=0.1$, 0.08, and 0.05. 
Here, panels (a) and (b) plot the isolines of  $(\hat{u}_x/u_\infty)/\kappa^{1/2}$ and $(\hat{u}_r/u_\infty)/\kappa^{1/2}$, respectively, using stretched coordinates $x/\kappa$ and $(r-1)/\kappa$, and the contour lines for $\kappa=0.1$, 0.08, and 0.05 are superimposed. This means that the plot region is shrinking to the edge as $\kappa$ becomes small in the original unstretched coordinate $(x,r)$.
It is seen that the isolines for three different values of $\kappa$ almost overlap both for $\hat{u}_x$ and $\hat{u}_r$. A similar structure is observed in the temperature and pressure fields, as shown in Fig.~\ref{fig:scale_tau_mod_ol_ver2}, where the isolines of $(\tau/u_\infty)/\kappa^{1/2}$ and $(P/u_\infty)/\kappa^{1/2}$ are plotted using the same stretched coordinates.
These observations indicate the following property of the flow field near the edge when $\kappa$ is small. That is, the deviation of the flow velocity from the corresponding Stokes value and the pressure and temperature have self-similar structures near the edge, which can be expressed in the form $\kappa^{1/2}f(x/\kappa,(r-1)/\kappa)$, where $f$ denotes $\hat{u}_x$, $\hat{u}_r$, $\tau$, or $P$. On the other hand, in the bulk region, the deviation of the flow velocity $(\hat{u}_x,\hat{u}_r)$, pressure $P$, and temperature $\tau$ decays in proportion to $\kappa$ as $\kappa \to 0$. This difference in the asymptotic behavior between the near-edge and bulk regions can be more clearly seen in Fig.~\ref{fig:macro_decay}, where the decay properties of $h= \hat{u}_r$, $\hat{u}_x$, $P$, and $\tau$ with respect to $\kappa$ are compared from two different viewpoints. That is, in panels (a) and (b), $h(x,r)$ at $(x,r)=(0_+,1_-)=\lim_{\epsilon \downarrow 0}(0_+,1-\epsilon)$ and $(x,r)=(\kappa,1)$ are plotted in log-log scale, respectively, as a function of $\kappa$, while (c) and (d) show $h(x,r)$ at $(x,r)=(0.5,1)$ and $(0.5,0.5)$, respectively, as a function of $\kappa$. Note that $\hat{u}_x$ is not shown in (a) because the quantity is identically zero on the disk surface. In (a) and (b), $|h|$ decays in proportion to $\kappa^{1/2}$, being consistent with Figs.~\ref{fig:scaled_velocity} and \ref{fig:scale_tau_mod_ol_ver2}. On the other hand, in (c) and (d), $|\hat{u}_x|$, $|\hat{u}_r|$, and $|P|$ decay in proportion to $\kappa$. Regarding $|\tau|$, the plot exhibits a slight deviation from the trend $\propto \kappa$ in (d) for $\kappa \le 0.03$. However, this deviation is attributed to numerical errors, as seen from the results using coarser grids (with $N_{\xi}$ reduced to two-thirds and $N_\eta$ halved), which are overplotted in the figures using open symbols. Note that the decay of $\tau$ appears slightly faster in (d). Based on additional data not shown here, we anticipate that $\tau$ also follows the same $\propto \kappa$ trend as $\kappa$ tends to zero, although further numerical studies are needed to confirm this conclusively. In summary, these results support the appearance of a kinetic boundary layer localized near the edge, and that the fluid variables in this region can be described in a self-similar manner, as mentioned previously.

\subsection{\label{subsec:thermal_polarization}Thermal polarization in the near-continuum regime}

We now turn our attention to the thermal polarization effects observed in the near-continuum regime. As previously noted, the gas temperature is lower on the downstream side and higher on the upstream side of the disk, with this temperature inhomogeneity becoming more pronounced near the disk edge than at its center as the Knudsen number decreases. To better understand this behavior, it is instructive to draw an analogy with the thermal edge flow, a flow induced near the edge of a uniformly heated plate \citep{Sone07,Aoki-Sone-Masukawa95,Sone-Yoshimoto97,Taguchi-Aoki_JFM12}.

In the case of a thermal edge flow, the sharp bending of isothermal lines is the key mechanism driving the flow. To illustrate this mechanism, we estimate the magnitude of the temperature-induced flow near the edge using the asymptotic theory \citep[the generalized slip-flow theory][]{Sone07}. Although this theory is not strictly valid near the edge, the theory is expected to capture the main physical mechanism and yield a reasonable estimate of the flow magnitude \citep{Sone-Yoshimoto97,Taguchi-Aoki_JFM12}. Consider a thin circular disk (radius $L$) immersed in a quiescent gas, whose surface temperature is maintained at $T_0 (1+\tau_{\rm w})$, where $T_0$ is a reference temperature and $\tau_{\rm w}$ is a small constant. We use the same notation as in the original problem. Assuming that the gas temperature $T_0(1+\tau)$ satisfies the Laplace equation, $\tau$ is expressed as $\tau = \frac{2}{\pi} \tau_{\rm w} \text{arccot}(\sinh \xi)$, where the same oblate spherical coordinates as in \eqref{eq:oblate} are used; in this coordinate system, the disk corresponds to $\xi=0$, which is equivalent to $x_1=0$ and $x_2^2 + x_3^2 < 1$ (or $X_1=0$ and $X_2^2+X_3^2 < L^2$ in dimensional form). Using the Taylor expansion, the temperature field near the edge can be approximated as
\begin{align} \label{e:tau_asymptotic}
    \frac{\tau}{\tau_{\rm w}} = 1 - \frac{2\sqrt{2}}{\pi} t^{1/2} \cos \left(\frac{\varphi}{2}\right) + O(t^{3/2}), \quad t \ll 1,
\end{align}
where $t = \sqrt{x^2+(r-1)^2}$ represents the distance from the edge in a meridian plane (say $x_3=0$) and $\varphi$ is the polar angle measured from the positive $x_2$-axis in that plane (see Fig.~\ref{fig:physical_estimate}(a); see also Sec.~S2 of Supplementary Materials for the derivation). Next, following \citet{Sone-Yoshimoto97,Sone07}, we relate the thermal edge flow to the thermal slip (also known as the thermal creep flow), which is a flow induced along a surface when the gas temperature varies tangentially along that surface. For the present situation, the slip velocity on the disk is given by 
\begin{align} \label{eq:thermal_slip}
    u_r = \hat{K}_1 \kappa \frac{\partial \tau}{\partial r} \quad (\text{on $x=0_\pm$}),
\end{align}
where $\hat{K}_1>0$ is the thermal-slip coefficient and $\kappa$ is the scaled mean free path (i.e., the Knudsen number) defined by \eqref{eq:def_kappa}. It should be noted that Eq.~\eqref{eq:thermal_slip} is derived under the assumption that the solution to the Boltzmann equation varies moderately, except within the Knudsen layer \citep{Sone07}. Therefore, it cannot be directly applied to flow quantities in the vicinity of the edge, where variations occur on the scale of the mean free path even in the tangential direction along the surface. However, the essential point in this argument is the temperature difference over the scale of a mean free path, implying that molecules arriving at a point on the disk from nearby locations can exhibit different average molecular speeds. This can be seen by rewriting the gradient term as $ \frac{\partial \tau}{\partial r}(x,r) \sim (\tau(x,r+\kappa/2)-\tau(x,r-\kappa/2))/\kappa$. In this sense, \eqref{eq:thermal_slip} is qualitatively valid in the vicinity of the edge, although the precise value of $\hat{K}_1$ does not make sense. To estimate  $\partial \tau/\partial r$ in the vicinity of the edge, we consider the point P located at $(x,r)=(\frac{\sqrt{3}}{2}\kappa,1)$, and evaluate the temperature gradient using two neighboring points separated by one mean free path: A: $(x,r)=(\frac{\sqrt{3}}{2}\kappa,1+\frac{1}{2}\kappa)$ and B: $(x,r)=(\frac{\sqrt{3}}{2}\kappa,1-\frac{1}{2}\kappa)$. The derivative at P can then be approximated as $\partial \tau/\partial r \sim (\tau(\mathrm{A}) - \tau(\mathrm{B}))/\kappa \sim -\tau_{\rm w}\sqrt{2} (\sqrt{3}-1) /\kappa^{1/2}$. Hence, \eqref{eq:thermal_slip} gives the slip velocity near the edge $u_r \sim - \tau_{\rm w}\hat{K}_1 \sqrt{2}(\sqrt{3}-1) \kappa^{1/2}$. This shows that the magnitude of the thermal edge flow around the disk scales as $\kappa^{1/2}$, or equivalently $\ell^{1/2}$ with $\ell$ being the mean free path. It should be noted that, although the $\kappa^{1/2}$ scaling has not been directly verified by solving the thermal edge flow around a disk, a similar asymptotic estimate \citep{Sone-Yoshimoto97} and its numerical confirmation for a two-dimensional plate \citep{Taguchi-Aoki_JFM12} support the validity of this scaling law in the present disk geometry.

\begin{figure}
\centering
    \subfigure[]{\includegraphics[scale=1.1]{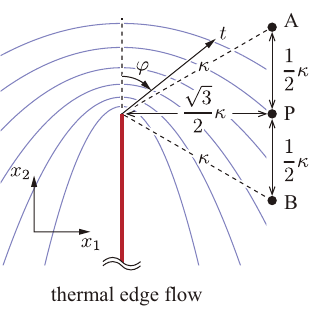}}
    \qquad
    \subfigure[]{\includegraphics[scale=1.1]{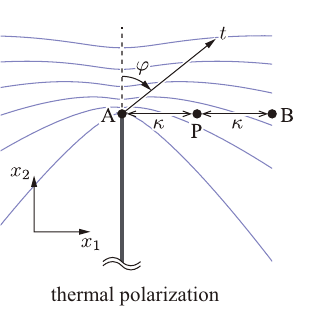}}
    \caption{Polar coordinates $(t,\varphi)$ and the location of the points P, A, and B near the disk edge in the plane $x_3=0$. The solid segment represents the disk and $\kappa$ the scaled mean free path (see \eqref{eq:def_kappa}). Isolines of the temperature $\tau$ and the flow-velocity component $u_x$ around the edge are schematically shown in (a) and (b), respectively.}
    \label{fig:physical_estimate}
\end{figure}

We now return to the original problem of uniform flow past a disk and apply a similar argument to estimate the magnitude of the temperature variation near the edge. Following \citet{Aoki-Sone-PHF-1987}, a key concept here is the temperature jump at the surface of a body. Typically, a temperature jump arises when the normal component of the heat flux does not vanish on the boundary. However, in the present setting, where the overall temperature variation is weak, this contribution is negligibly small. Instead, we focus on a second-order temperature jump, which appears when the quantity $n_in_jn_k \partial/\partial x_k(\partial u_i/\partial x_j + \partial u_j/\partial x_i)$ does not vanish at the surface \citep{Sone07}, where $n_i$ denotes the unit normal vector pointing into the gas. 
This contribution
constitutes the second-order jump condition in the asymptotic theory. 
While typically small, it can be significant near the disk edge due to the amplification of velocity gradients in that region.

In the present notation, the aforementioned temperature jump on the disk can be written as
\begin{align} \label{eq:temperature-jump}
    \tau = \pm 2d_4 \kappa^2 \frac{\partial^2 u_x}{\partial x^2} \quad (\text{on $x=0_\pm$}),
\end{align}
where $d_4 > 0$ is the associated temperature-jump coefficient, and the factor $\kappa^2$ reflects that this condition arises at the second order in the asymptotic expansion. Again, this condition is strictly valid when the spatial variation of fluid quantities is moderate, and thus it should be interpreted as a qualitatively accurate relation in the vicinity of the edge.
To estimate the right-hand side, we perform a Taylor expansion of the Stokes velocity near the disk edge, yielding for $t \ll1$,
\begin{subequations} \label{e:velo_asymptotic}
\begin{align}
 u_x/u_\infty & = \frac{2\sqrt{2}}{\pi} t^{1/2} \cos^3 \left(\frac{\varphi}{2}\right) (1 + O(t)),
 \label{e:ux_edge}
 \\
 u_r/u_\infty & = -\frac{2\sqrt{2}}{\pi} t^{1/2} \cos^2\left(\frac{\varphi}{2} \right) \sin \left(\frac{\varphi}{2} \right) (1 + O(t)),
\end{align}  
\end{subequations}
where $t$ and $\varphi$ are defined as before (see Sec.~S2 of Supplementary Materials for the derivation). This time, we consider the point P: $(x,r)=(\kappa,1)$ in the vicinity of the edge, and its neighboring points A: $(x,r)=(0_+,1)=\lim_{\epsilon \downarrow 0}(0+\epsilon,1)$ and B: $(x,r)=(2\kappa,1)$ (see Fig.~\ref{fig:physical_estimate}(b)). The second derivative $\partial^2 u_x/\partial x^2$ at point P can be estimated using a finite-difference approximation:
$\partial_x^2 u_x \sim (u_x(\mathrm{A}) - 2u_x(\mathrm{P}) + u_x(\mathrm{B}))/\kappa^2=- u_\infty (\sqrt{2}/\pi)(\sqrt{2}-1) \kappa^{-3/2}$. Substituting this into the jump condition \eqref{eq:temperature-jump} yields a negative temperature jump at point P, i.e., $\tau/u_\infty \sim - 2 (\sqrt{2}/\pi)(\sqrt{2}-1) d_4 \kappa^{1/2}$. Similarly, at the symmetric point Q: $(x,r)=(-\kappa,1)$, we obtain a positive temperature jump, $\tau/u_\infty \sim 2 (\sqrt{2}/\pi)(\sqrt{2}-1) d_4 \kappa^{1/2}$. This asymptotic scaling of the temperature variation with the power $1/2$ is consistent with our numerical results. In this way, the observed thermal polarization in the near-continuum regime can be reasonably interpreted as a manifestation of the second-order temperature jump effect localized near the edge.

Finally, it should be noted that the appearance of the same power $1/2$ in both the thermal edge flow and the thermal polarization in the uniform flow problem is not coincidental. Rather, it stems from the fact that both the Stokes and Laplace equations admit the same leading-order singularity near a corner. In this sense, the magnitude of the edge effect is governed purely by geometrical considerations.

\begin{figure*}[tbp]
\begin{center}
\includegraphics[width=0.6\linewidth]{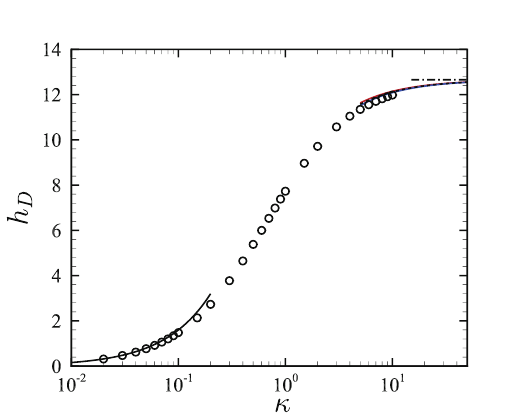}
\caption{The dimensionless force $h_D$ vs $\kappa$.
The symbol $\circ$ represents the present numerical results.
The solid curve represents $h_D=16 \gamma_1 \kappa$ with $\gamma_1 = 1$, corresponding to the Stokes equation with no-slip boundary conditions. The dash-dot line indicates the value in the free-molecular limit, given by $h_D(\infty) = \sqrt{\pi}(\pi+4)$. The dashed curves (black, red, and blue) include a $\kappa^{-1}$-order correction term to the free-molecular solution in the case of a hard-sphere gas \citep{Sengers_etal_2014}. Note that \citet{Sengers_etal_2014} reports the coefficients with some numerical uncertainty. The blue and red curves in the figure represent the upper and lower bounds, respectively, corresponding to the range of uncertainty in the reported coefficient.
}
\label{fig:drag}
\end{center}
\end{figure*}

\begin{table}
\begin{center}
\caption{The dimensionless drag force $h_D$ as a function of $\kappa = (\sqrt{\pi}/{2}) \mathrm{Kn}$ for typical values of $\kappa$. The values of $c_1$ are also shown. Here, the corresponding values of $\kn$ are included for reference, and the values shown in parentheses represent those computed from $c_1$ using the relation \eqref{eq:relation_hd_c1}. A more complete dataset, including results for other $\kappa$ values, is provided in Table~S1 of Supplementary Materials, where the values of $c_2$ and $c_3$ are also included.}
\label{tab:drag2}
\vspace{\baselineskip}
\begin{tabular}{lrrrrclrrrr} \hline\hline
\multicolumn{1}{c}{$\kappa$} & \multicolumn{1}{c}{$\,\, \mathrm{Kn}$} & 
\multicolumn{2}{c}{$h_D$} & \multicolumn{1}{c}{$c_1$} & \hspace{0.5cm} &
\multicolumn{1}{c}{$\kappa$} & \multicolumn{1}{c}{$\,\, \mathrm{Kn}$} & 
\multicolumn{2}{c}{$h_D$} & \multicolumn{1}{c}{$c_1$} \\ \hline
$0.02$ & $\,\, 0.0226$ & $0.3160$ & ($0.3115$) & $0.6198$ & \hspace{0.5cm} &
$1$ & $\,\, 1.1284$ & $7.7308$ & ($7.7298$) & $0.3076$ \\
$0.05$ & $\,\, 0.0564$ & $0.7710$ & ($0.7645$) & $0.6084$ & \hspace{0.5cm} &
$2$ & $\,\, 2.2568$ & $9.7148$ & ($9.7237$) & $0.1934$ \\
$0.1$ & $\,\, 0.1128$ & $1.4815$ & ($1.4762$) & $0.5873$ & \hspace{0.5cm} &
$5$ & $\,\, 5.6419$ & $11.3465$ & ($11.3616$) & $0.0904$ \\
$0.2$ & $\,\, 0.2257$ & $2.7320$ & ($2.7277$) & $0.5427$ & \hspace{0.5cm} &
$10$ & $\,\, 11.2838$ & $11.9809$ & ($11.9552$) & $0.0476$ \\
$0.5$ & $\,\, 0.5642$ & $5.3814$ & ($5.3803$) & $0.4282$ & \hspace{0.5cm} &
 & & & & \\ \hline\hline
\end{tabular}
\end{center}
\end{table}

\section{\label{sec:force}Force acting on the disk}

Finally, we present the numerical results for the total force acting on the disk.
Figure~\ref{fig:drag} shows $h_D$ as a function of $\kappa$ with some representative numerical values tabulated in Table~\ref{tab:drag2}; a more complete table is provided in Table~S1 of Supplementary Materials. The $h_D$ increases monotonically in $\kappa$ and tends to approach the free molecular value $h_D(\infty)=\sqrt{\pi}(\pi+4) \approx 12.66$ as $\kappa \to \infty$ (dash-dot line; see Eq.~\eqref{eq:h_D_fmg}).
A $\kappa^{-1}$-order correction to the free molecular solution has been obtained in  \citet{Sengers_etal_2014} for a hard-sphere gas [Eq.~(5.16) of the reference]. The results are shown by the dashed curves in the figure for comparison. The present study agrees with the theoretical prediction. Note that, to mitigate model inconsistency in the comparison, we applied the conversion $1.270042427 \ell_0^{\text{HS}} = \ell_0^{\text{BGK}}$ before plotting the result from \citet{Sengers_etal_2014}, where $\ell_0^{\text{HS}}$ and $\ell_0^{\text{BGK}}$ stand for the mean free path of the hard-sphere gas and the BGK model, respectively. This relation is derived by equating the viscosity for the two models. It should also be mentioned that the coefficients reported in \citet{Sengers_etal_2014} include some numerical uncertainty. In the figure, the blue and red curves represent the upper and lower bounds, respectively, corresponding to the reported uncertainty in the coefficients.
If the flow past a circular disk is considered based on the Stokes equation with no-slip boundary conditions, the force exerted on the disk is expressed as $F=16UL \mu_0$ \citep{Payne_Pell_1960,Happel+Brenner_1983}. 
The corresponding value of $h_D$ for this case [see Eq.~\eqref{eq:drag_force} and the sentence following \eqref{eq:far_asy}] is $h_D=16 \gamma_1 \kappa$, where $\gamma_1=1$ for the BGK model, which is shown by the solid curve in the figure as a limiting case for $\kappa \to 0$.
The numerical results for $h_D$ tend to approach this value as $\kappa$ is decreased, as seen from Fig.~\ref{fig:drag}. 

In Table~\ref{tab:drag2}, the computed values of $c_1$
appearing in \eqref{eq:far_asy} are also presented as functions of $\kappa$ (and $\kn$); the values of $c_2$ and $c_3$ are presented in Table~S1 of Supplementary Materials.
The values of $h_D$ obtained from $c_1$ via the relation \eqref{eq:relation_hd_c1} are shown in parentheses. Ideally, these should coincide with those directly calculated using \eqref{eq:def_h_D}; however, due to numerical errors, slight discrepancies arise. In general, computing $c_1$ is more demanding than computing $h_D$ from \eqref{eq:def_h_D}, as $c_1$ is determined through matching the solution in the far-field region, where the perturbation becomes small. Thus, the agreement between these two values serves as an indicator of the accuracy of the present computation. As shown in the table, the overall agreement is good, although deviations become more pronounced at both small and large Knudsen numbers.

\section{\label{sec:conclusion}Concluding remarks}
In this study, we have investigated the steady flow of a rarefied gas past a circular disk based on the BGK model of the Boltzmann equation. Although this problem is classical in kinetic theory, the edge effect in the low-speed limit has not been previously addressed, and accurately resolving the velocity distribution function in a three-dimensional axisymmetric flow poses a nontrivial numerical challenge.
The main contributions of the study are summarized as follows:
\begin{enumerate}
    \item We have shown that accurate numerical solutions for axisymmetric rarefied gas flows, while fully preserving the discontinuity structure of the velocity distribution function, are now within reach via an integral-equation approach. The method enables us to elucidate the behavior of the macroscopic quantities in the vicinity of the edge, where the edge layer exhibits a $\kappa^{1/2}$ (or equivalently, $\mathrm{Kn}^{1/2}$) scaling.
    \item Thermal polarization has been observed over the disk. In the near-continuum regime, this effect can be qualitatively interpreted in terms of the temperature jump in the asymptotic theory of the Boltzmann equation, although the applicability of this theory to the present configuration is not fully ensured. In addition, the magnitudes of thermal polarization and thermal edge flow exhibit the same asymptotic scaling with respect to the Knudsen number as it approaches zero. This similarity is attributed to a common geometrical origin.
    \item We obtained the drag force acting on the disk for a wide range of the Knudsen number. As $\mathrm{Kn} \to 0$, the computed force converges to the value predicted by the Stokes equation with no-slip boundary conditions. Our results reasonably agree with the existing results for the near-free-molecular regime.
\end{enumerate}

Finally, we mention several possible directions for future work. One natural extension is to consider the case where the disk surface temperature is nonuniform. For example, if the two sides of the disk are maintained at different uniform temperatures, the system can serve as a model for a self-propelled particle or as an idealized vane of a Crookes radiometer. The flow field in such a configuration could be analyzed using the numerical approach developed in this study. Another interesting direction is to develop a theoretical framework that explains the asymptotic behavior of the flow as $\mathrm{Kn} \to 0$. It is known that the shear stress vanishes on the disk surface when the Stokes solution is assumed, and consequently, the conventional Navier slip condition fails to provide a correction to the Stokes flow \citep{Sherwood_PhysFluids_2012}. Identifying the appropriate correction term would represent an important next step toward a more complete understanding of the flow behavior in the near-continuum regime.

\begin{acknowledgments}
This work was supported by JSPS KAKENHI Grant Numbers JP22K03924, JP22K18770, JP25K01156 and JST SPRING Grant Number JPMJSP2110.
Numerical computations were carried out using the supercomputer of ACCMS, Kyoto University.
\end{acknowledgments}

\vspace{\baselineskip}
\textbf{Declaration of Interest.} The authors report no conflict of interest.

\appendix

\section{\label{app:matching}Process of numerical matching}

In this appendix, we explain the detailed process of matching \eqref{eq:far_asy} with the numerical solution.
First, expanding \eqref{eq:far_asy} in terms of the inverse power of $\chi = e^{\xi}$, we obtain
\begin{subequations}
\label{eq:far_asy_order}
\begin{align}
    \omega/u_{\infty} & = -\frac{4 \left(2 c_1 \kappa +c_3\right) \cos \eta}{\chi^2} + O(\chi^{-4}), \label{eq:omega_far_order}
    \\
    u_{x}/u_{\infty} & = 1 - \frac{2 c_1 (1 + \cos^2 \eta)}{\chi} + 
    \frac{c_1 (4 + 7 \cos 2 \eta + \cos 4 \eta) - 8 c_2 (\cos 2 \eta + \frac{1}{3})}{\chi^3} + O(\chi^5),
    \label{eq:ux_far_order}
    \\
    u_r/u_{\infty} & = \sin 2 \eta \left[-\frac{c_1 }{\chi} + 
    \frac{c_1 (2 \cos 2 \eta +7)-8 c_2}{\chi^3} + O(\chi^{-5})\right],
    \label{eq:ur_far_order}
    \\
    \tau/u_{\infty} & = \frac{4 c_3 \cos \eta}{ \chi^2} + O(\chi^{-4}),
    \label{eq:tau_far_order}
    \\
    P/u_\infty & = -\frac{8c_1 \kappa \cos \eta}{\chi^2} + O(\chi^{-4}).
    \label{eq:P_far_order}
\end{align}
\end{subequations}
Note that $\chi$ is proportional to $\hat{r}=(x^2+r^2)^{1/2}$ when $\hat{r} \gg 1$, 
since $x \to \frac{\chi}{2} \cos \hat{\theta}$, $r \to \frac{\chi}{2} \sin \hat{\theta}$ as $\xi \to \infty$, where $\hat{\theta}= \tan^{-1}(r/x)$. 
In general, when numerically solving a boundary-value problem posed in an unbounded domain, numerical errors arising from the truncation of the domain and the numerical integration are a significant concern. To mitigate this issue, we solve for $\phi'=\phi-\phi_\text{asy}$ instead of solving for $\phi$,
where $\phi_\text{asy}$ represents the asymptotic solution that approximates the behavior of $\phi$ in the far-field region and defined by
\begin{align}
    \phi_\text{asy}/u_\infty = 2\zeta_x - \frac{2c_1}{\chi}\left[2(1+\cos^2\eta)\zeta_x + \zeta_r \sin 2\eta\right].
\end{align}
Note that $\phi_\text{asy}$ corresponds to the $\chi^{-1}$-order correction to the uniform equilibrium distribution at infinity. Since $\mathcal{L}(\phi_\text{asy})=0$, the function $\phi'$ satisfies the equation $\zeta_i \partial \phi'/\partial x_i = \frac{1}{\kappa}\mathcal{L}(\phi') - \zeta_i \partial \phi_\text{asy}/\partial x_i$. The integral equation for $\phi'$ is derived in the same manner as for $\phi$, and the resulting equation, which we omit here, is solved as outlined in the main text. This approach helps in improving the accuracy of the numerical solution, in particular in regions far from the disk, where the asymptotic form dominates.

Let $P_{\text{asy}}^*/u_\infty=-8c_1 \kappa \cos \eta/e^{2\xi}$, $\tau_{\text{asy}}^* = 4 c_3 \cos \eta/e^{2\xi}$, and $u_{r,\text{asy}}^*/u_\infty = [c_1 (2 \cos 2 \eta +7)-8 c_2 ] \sin 2 \eta/e^{3\xi}$.
Here, $P_{\text{asy}}^*$ and $\tau_{\text{asy}}^*$ correspond to the leading-order terms from \eqref{eq:P_far_order} and \eqref{eq:tau_far_order}, while $u_{r,\text{asy}}^*$ represents the $\chi^{-3}$-order term from \eqref{eq:ur_far_order}.
We also set $u_{r,\text{asy}}^* = c_1 u_{r,\text{asy}}^{*1} + c_2 u_{r,\text{asy}}^{*2}$, where $u_{r,\text{asy}}^{*1} = (2 \cos 2 \eta +7) \sin 2 \eta/e^{3\xi}$ and $u_{r,\text{asy}}^{*2} = -8 \sin 2 \eta/e^{3\xi}$. 
Note that these are functions of $(\xi,\eta)$, e.g., $P_{\text{asy}}^* = P_{\text{asy}}^*(\xi,\eta)$.

To determine $c_1$ and $c_3$, we note that for a given $\xi=\xi_0$, both
$P_{\text{asy}}^*(\xi_0,\eta)/u_\infty$ and $\tau_{\text{asy}}^*(\xi_0,\eta)/u_\infty$ are proportional to $\cos \eta$, with proportionality factors $-8c_1 \kappa/e^{2\xi_0}$ and $4c_3 /e^{2\xi_0}$, respectively. By fitting the numerical data for $P$ and $\tau$ on the curve $\xi = \xi_0$ with $P_{\text{asy}}^*(\xi_0,\eta)$ and $\tau_{\text{asy}}^*(\xi_0,\eta)$, we can determine $c_1$ and $c_3$. We have used the least square method for fitting.
To determine $c_2$, we first note that in our deviational formulation, the integral $\int \zeta_r \phi' E d\bm{\zeta}$ produces $u_r'$, which is related to $u_r$ through the relation $u_r'= u_r + c_1 \sin 2\eta/e^\xi$. Therefore, $u_r'$ approaches $u_{r,\text{asy}}^*(\xi,\eta)$ as $\xi \to \infty$. Based on this consideration, we fit the numerical data for $u_r' - c_1 u_{r,\text{asy}}^{*1}(\xi_0,\eta)$ at $\xi=\xi_0$ with $c_2 u_{r,\text{asy}}^{*2}(\xi_0,\eta)$ by the least square method to determine $c_2$. Note that in this process, we use already determined $c_1$.

This process is executed at the end of each iteration in our numerical calculations and is repeated until convergence of both $\phi'$ and $c_i$ is achieved.
The choice of $\xi_0$ depends on $\kappa$; for example, $\xi_0=2.14$ ($\xi_{\max} = 3.1$) for $\kappa=0.05$, $\xi_0=2.54$ ($\xi_{\max} = 3.5$) for $\kappa=0.1$, $\xi_0=3.48$ ($\xi_{\max} = 4.5$) for $\kappa=0.5$, $\xi_0=3.87$ ($\xi_{\max} = 5$) for $\kappa=1$, and $\xi_0=4.89$ ($\xi_{\max} = 6$) for $\kappa=5$.
In most cases, the update of $c_i$ must be controlled using under-relaxation to ensure convergence. The computed values of $c_1$, $c_2$, and $c_3$ are summarized in Table~\ref{tab:drag2}.

\section{\label{app:accuracy}Accuracy of the numerical analysis}

In this appendix, we first summarize the lattice system and then present the results of various accuracy tests.

We begin by summarizing the lattice system used in the present numerical computations.
According to \eqref{eq:config_xr_lp_zeta}, the lattice points for the spatial variables $\xi$ and $\eta$ are defined by the following functions:
\begin{align}
\label{eq:def_g_xi_g_eta}
    g_{\xi}(i) = \begin{cases}
        \displaystyle \xi_{\max} \frac{i}{N_{\xi}} &(\kappa \geq 0.15), \\
        \displaystyle \xi_{\max} \left( \frac{i}{N_{\xi}} \right)^2 &(\kappa \leq 0.1),
    \end{cases} \quad
    g_{\eta}(j) = \begin{cases}
        \displaystyle \frac{\pi}{2} \frac{j}{N_{\eta}} &(\kappa \geq 0.15), \\
        \displaystyle \frac{\pi}{2} \frac{j}{N_{\eta}} \left( 2 - \frac{j}{N_{\eta}} \right) &(\kappa \leq 0.1).
    \end{cases}
\end{align}
Here, $N_{\xi} = 195$ and $N_{\eta} = 64$ are used for all values of $\kappa$, while $\xi_{\max}$ is appropriately chosen depending on $\kappa$.
Typical values of $\xi_{\max}$ are listed in the last paragraph of Appendix~\ref{app:matching}.
For the molecular velocity variables $\zeta$, $\theta_{\zeta}$, and $\varphi_{\zeta}$, we omit the explicit forms of $g_{\zeta}(y)$, $g_{\theta_{\zeta}}(y)$, $g_{\theta_{\zeta}}^{-}(y)$, $g_{\theta_{\zeta}}^{+}(y)$, $g_{\theta_{\zeta}}^{\flat}(y)$, $g_{\theta_{\zeta}}^{\natural}(y)$, $g_{\theta_{\zeta}}^{\sharp}(y)$, $g_{\varphi_{\zeta}}(y)$, $g_{\varphi_{\zeta}}^{-}(y)$, and $g_{\varphi_{\zeta}}^{+}(y)$ appearing in \eqref{eq:def_c_zetak}, \eqref{eq:def_phim_thetal_0r1}, and \eqref{eq:def_phim_thetal_r1}. These are linearly increasing, except for $g_{\zeta}(y)$ and $g_{\varphi_{\zeta}}^{-}(y)$, which are quadratic to ensure denser lattice point distributions near $\zeta=0$ and $\varphi_{\zeta} = \varphi_{\zeta *}$, respectively. The values of $\zeta_{\max}$, $N_{\zeta}$, and $N_{\theta_{\zeta}}$ are fixed for all values of $\kappa$: $\zeta_{\max}=5$, $N_{\zeta}=32$, and $N_{\theta_{\zeta}}=64$. In contrast, $N_{\varphi_{\zeta}}$ varies with $\kappa$ as follows: $N_{\varphi_{\zeta}} = 128$ ($\kappa \leq 1.5$), $256$ ($2 \leq \kappa \leq 4$), $512$ ($5 \leq \kappa \leq 7$), and $1024$ ($\kappa \geq 8$). 

We now comment on the lattice system used along characteristic curves for evaluating the integral in \eqref{eq:update_rule}. In the numerical computations, the integral interval $[0,s_{*,ijlm}]$ is truncated at $125\kappa$ if the length of the backward characteristic curve exceeds this value. The (truncated) interval is then divided into subintervals for application of the four-point Gauss--Legendre quadrature formula (see \ref{subsec:outline}). The number of subintervals is proportional to the length of the backward characteristic curve, ranging from $1$ to $64$. To accurately capture the variation of the integrand, the subinterval lengths are adapted such that the lattice points are concentrated near the starting point of the backward characteristic curve and in regions where the curve passes close to the disk edge. 

To verify the accuracy of the numerical results, various numerical tests were conducted. 
We refer to the parameter configuration described above as the ``standard setting." In each test, we select a subset of variables from $\{ \xi, \eta, \zeta, \theta_{\zeta}, \varphi_{\zeta} \}$ and double the number of lattice points for the chosen variables, while keeping all the other parameters unchanged. Throughout this appendix, superscripts are used to indicate the parameter setting under which a quantity is computed: the standard setting is denoted by the superscript $(\text{sta})$, while a refined lattice system is identified by listing the modified variables in parentheses in place of ``$\text{sta}$."
It should be noted that the other parameters, such as $\xi_{\max}$, $\zeta_{\max}$, and $\xi_0$, and the forms of the functions used in \eqref{eq:config_xr_lp_zeta}, \eqref{eq:def_c_zetak}, \eqref{eq:def_phim_thetal_0r1}, and \eqref{eq:def_phim_thetal_r1}, are kept fixed throughout these tests.

\begin{table}
\begin{center}
\caption{Values of $\Delta_{\mathcal{F}}^{(\xi,\eta)}$ and $\Delta_{h}^{(\xi,\eta)}$ for $\kappa = 5$, $0.5$, and $0.05$
($\mathcal{F} = h_D, c_1, c_2, c_3$, $h = \omega, u_x, u_r, \tau$).}
\label{tab:acc_xieta}
\begin{tabular}{lcccc} \hline
    &\qquad $\Delta_{h_D}^{(\xi,\eta)}$ &\quad $\Delta_{c_1}^{(\xi,\eta)}$ 
    &\quad $\Delta_{c_2}^{(\xi,\eta)}$ &\quad $\Delta_{c_3}^{(\xi,\eta)}$ \\ \hline
    $\kappa=5$ &\qquad $2.3 \times 10^{-8}$ &\quad $2.0 \times 10^{-5}$ 
    &\quad $1.3 \times 10^{-3}$ &\quad $5.0 \times 10^{-4}$ \\
    $\kappa=0.5$ &\qquad $4.3 \times 10^{-7}$ &\quad $8.7 \times 10^{-6}$ 
    &\quad $1.6 \times 10^{-4}$ &\quad $1.1 \times 10^{-4}$ \\
    $\kappa=0.05$ &\qquad $2.7 \times 10^{-5}$ &\quad $2.0 \times 10^{-4}$ 
    &\quad $2.7 \times 10^{-3}$ &\quad $2.0 \times 10^{-1}$ \\ \hline
    &\qquad $\Delta_{\omega}^{(\xi,\eta)}$ &\quad  $\Delta_{u_x}^{(\xi,\eta)}$ 
    &\quad $\Delta_{u_r}^{(\xi,\eta)}$ &\quad  $\Delta_{\tau}^{(\xi,\eta)}$ \\ \hline
    $\kappa=5$ &\qquad $6.0 \times 10^{-5}$ &\quad $8.0 \times 10^{-5}$ 
    &\quad $4.1 \times 10^{-5}$ &\quad $9.7 \times 10^{-6}$ \\
    $\kappa=0.5$ &\qquad $9.5 \times 10^{-5}$ &\quad $4.3 \times 10^{-5}$ 
    &\quad $5.9 \times 10^{-5}$ &\quad $1.8 \times 10^{-4}$ \\
    $\kappa=0.05$ &\qquad $4.6 \times 10^{-4}$ &\quad $2.9 \times 10^{-5}$ 
    &\quad $5.4 \times 10^{-5}$ &\quad $4.5 \times 10^{-3}$ \\ \hline
\end{tabular}
\end{center}
\end{table}

\begin{table}
\begin{center}
\caption{Values of $\Delta_{\mathcal{F}}^{(\alpha)}$ and $\Delta_{h}^{(\alpha)}$ for $\kappa = 5$, $0.5$, and $0.05$
($\alpha = \zeta, \theta_{\zeta}, \varphi_{\zeta}$, 
$\mathcal{F} = h_D, c_1, c_2, c_3$, $h = \omega, u_x, u_r, \tau$).}
\label{tab:acc_mvv}
\begin{tabular}{lcccc} \hline
    &\qquad $\Delta_{h_D}^{(\zeta)}$ &\quad $\Delta_{c_1}^{(\zeta)}$ 
    &\quad $\Delta_{c_2}^{(\zeta)}$ &\quad $\Delta_{c_3}^{(\zeta)}$ \\ \hline
    $\kappa=5$ &\qquad $1.2 \times 10^{-8}$ &\quad $3.4 \times 10^{-8}$ 
    &\quad $1.4 \times 10^{-5}$ &\quad $4.0 \times 10^{-5}$ \\
    $\kappa=0.5$ &\qquad $4.9 \times 10^{-8}$ &\quad $2.1 \times 10^{-7}$ 
    &\quad $3.8 \times 10^{-5}$ &\quad $2.2 \times 10^{-4}$ \\
    $\kappa=0.05$ &\qquad $1.1 \times 10^{-5}$ &\quad $2.0 \times 10^{-6}$ 
    &\quad $4.8 \times 10^{-6}$ &\quad $5.9 \times 10^{-4}$ \\ \hline
    &\qquad $\Delta_{h_D}^{(\theta_{\zeta})}$ &\quad $\Delta_{c_1}^{(\theta_{\zeta})}$ 
    &\quad $\Delta_{c_2}^{(\theta_{\zeta})}$ &\quad $\Delta_{c_3}^{(\theta_{\zeta})}$ \\ \hline
    $\kappa=5$ &\qquad $5.2 \times 10^{-6}$ &\quad $5.0 \times 10^{-4}$ 
    &\quad $1.2 \times 10^{-4}$ &\quad $5.8 \times 10^{-3}$ \\
    $\kappa=0.5$ &\qquad $7.9 \times 10^{-6}$ &\quad $1.8 \times 10^{-5}$ 
    &\quad $2.6 \times 10^{-5}$ &\quad $3.2 \times 10^{-4}$ \\
    $\kappa=0.05$ &\qquad $3.6 \times 10^{-6}$ &\quad $1.2 \times 10^{-6}$ 
    &\quad $4.3 \times 10^{-7}$ &\quad $1.1 \times 10^{-4}$ \\ \hline
    &\qquad $\Delta_{h_D}^{(\varphi_{\zeta})}$ &\quad $\Delta_{c_1}^{(\varphi_{\zeta})}$ 
    &\quad $\Delta_{c_2}^{(\varphi_{\zeta})}$ &\quad $\Delta_{c_3}^{(\varphi_{\zeta})}$ \\ \hline
    $\kappa=5$ &\qquad $7.8 \times 10^{-6}$ &\quad $9.8 \times 10^{-4}$ 
    &\quad $2.9 \times 10^{-3}$ &\quad $2.0 \times 10^{-2}$ \\ 
    $\kappa=0.5$ &\qquad $2.7 \times 10^{-5}$ &\quad $7.5 \times 10^{-5}$ 
    &\quad $1.3 \times 10^{-4}$ &\quad $2.4 \times 10^{-3}$ \\ 
    $\kappa=0.05$ &\qquad $1.4 \times 10^{-6}$ &\quad $3.2 \times 10^{-7}$ 
    &\quad $9.6 \times 10^{-8}$ &\quad $2.8 \times 10^{-5}$ \\ \hline
    &\qquad $\Delta_{\omega}^{(\zeta)}$ &\quad  $\Delta_{u_x}^{(\zeta)}$ 
    &\quad $\Delta_{u_r}^{(\zeta)}$ &\quad  $\Delta_{\tau}^{(\zeta)}$ \\ \hline
    $\kappa=5$ &\qquad $2.6 \times 10^{-8}$ &\quad $4.9 \times 10^{-8}$ 
    &\quad $4.1 \times 10^{-8}$ &\quad $1.2 \times 10^{-7}$ \\
    $\kappa=0.5$ &\qquad $1.5 \times 10^{-7}$ &\quad $5.2 \times 10^{-8}$ 
    &\quad $1.4 \times 10^{-7}$ &\quad $1.5 \times 10^{-6}$ \\
    $\kappa=0.05$ &\qquad $7.8 \times 10^{-6}$ &\quad $1.4 \times 10^{-6}$ 
    &\quad $4.0 \times 10^{-6}$ &\quad $6.0 \times 10^{-6}$ \\ \hline
    &\qquad $\Delta_{\omega}^{(\theta_{\zeta})}$ &\quad  $\Delta_{u_x}^{(\theta_{\zeta})}$ 
    &\quad $\Delta_{u_r}^{(\theta_{\zeta})}$ &\quad  $\Delta_{\tau}^{(\theta_{\zeta})}$ \\ \hline
    $\kappa=5$ &\qquad $3.3 \times 10^{-5}$ &\quad $1.3 \times 10^{-5}$ 
    &\quad $8.7 \times 10^{-5}$ &\quad $1.3 \times 10^{-4}$ \\
    $\kappa=0.5$ &\qquad $1.3 \times 10^{-4}$ &\quad $8.4 \times 10^{-6}$ 
    &\quad $3.8 \times 10^{-4}$ &\quad $2.1 \times 10^{-4}$ \\
    $\kappa=0.05$ &\qquad $2.0 \times 10^{-4}$ &\quad $2.7 \times 10^{-6}$ 
    &\quad $3.3 \times 10^{-4}$ &\quad $4.1 \times 10^{-4}$ \\ \hline
    &\qquad $\Delta_{\omega}^{(\varphi_{\zeta})}$ &\quad  $\Delta_{u_x}^{(\varphi_{\zeta})}$ 
    &\quad $\Delta_{u_r}^{(\varphi_{\zeta})}$ &\quad  $\Delta_{\tau}^{(\varphi_{\zeta})}$ \\ \hline
    $\kappa=5$ &\qquad $8.1 \times 10^{-6}$ &\quad $7.8 \times 10^{-6}$ 
    &\quad $8.2 \times 10^{-6}$ &\quad $7.8 \times 10^{-6}$ \\
    $\kappa=0.5$ &\qquad $2.9 \times 10^{-5}$ &\quad $3.1 \times 10^{-5}$ 
    &\quad $2.8 \times 10^{-5}$ &\quad $4.6 \times 10^{-5}$ \\
    $\kappa=0.05$ &\qquad $2.1 \times 10^{-4}$ &\quad $7.0 \times 10^{-5}$ 
    &\quad $6.1 \times 10^{-6}$ &\quad $3.4 \times 10^{-4}$ \\ \hline
\end{tabular}
\end{center}
\end{table}

First, to examine the sensitivity to spatial resolution, we simultaneously double $N_{\xi}$ and $N_{\eta}$, while keeping $N_{\zeta}$, $N_{\theta_{\zeta}}$, and $N_{\varphi_{\zeta}}$ fixed at their values in the standard setting. The variation in the computed drag $h_D$ and the constants $c_i$ in \eqref{eq:far_asy} is then evaluated by
\begin{align}
\label{eq:def_Delta_F_xieta}
    \Delta^{(\xi,\eta)}_{\mathcal{F}} 
    = \frac{|\mathcal{F}^{(\xi,\eta)} - \mathcal{F}^{\text{(sta)}}|}{|\mathcal{F}^{\text{(sta)}}|} \quad
    (\mathcal{F} = h_D, c_1, c_2, c_3).
\end{align}
The corresponding variation in the computed macroscopic variables is measured by
\begin{align}
\label{eq:def_Delta_h_xieta}
    \Delta_h^{(\xi,\eta)} 
    = \frac{\max_{i,j} | h_{2i,2j}^{(\xi,\eta)} - h_{ij}^{(\text{sta})} |}{\max_{i,j} | h_{ij}^{(\text{sta})} | } 
    \quad (h = \omega, u_x, u_r, \tau).
\end{align}
These values are summarized in Table~\ref{tab:acc_xieta} for $\kappa = 5$, $0.5$, and $0.05$.

Next, we turn our attention to the molecular velocity variables. Each of $N_{\zeta}$, $N_{\theta_{\zeta}}$, and $N_{\varphi_{\zeta}}$ is doubled individually, while $N_{\xi}$ and $N_{\eta}$ remain fixed. In the same fashion as \eqref{eq:def_Delta_F_xieta} and \eqref{eq:def_Delta_h_xieta}, the effect of increasing the number of lattice points are quantified by
\begin{align}
\label{eq:def_Delta_F_mvv}
    \Delta^{(\alpha)}_{\mathcal{F}} 
    = \frac{|\mathcal{F}^{(\alpha)} - \mathcal{F}^{\text{(sta)}}|}{|\mathcal{F}^{\text{(sta)}}|} \quad
    (\alpha = \zeta, \theta_{\zeta}, \varphi_{\zeta}, \,\, \mathcal{F} = h_D, c_1, c_2, c_3),
\end{align}
and
\begin{align}
\label{eq:def_Delta_h_mvv}
    \Delta_h^{(\alpha)} 
    = \frac{\max_{i,j} | h_{ij}^{(\alpha)} - h_{ij}^{(\text{sta})} |}{\max_{i,j} | h_{ij}^{(\text{sta})} | } 
    \quad (\alpha = \zeta, \theta_{\zeta}, \varphi_{\zeta}, \,\, h = \omega, u_x, u_r, \tau).
\end{align}
The corresponding results are summarized in Table~\ref{tab:acc_mvv} for $\kappa = 5$, $0.5$, and $0.05$.
The data presented in Tables~\ref{tab:acc_xieta} and \ref{tab:acc_mvv} indicate that the numerical errors are small and do not affect the main conclusions of the study.

\bibliographystyle{jfm}
\bibliography{mydatabase_20251017}

\end{document}